\documentclass[12pt, preprint]{aastex}
 \usepackage{epsfig}
\def\deg{\ifmmode^\circ\else$^\circ$\fi}

\def\gs{{_>\atop^{\sim}}}

\def\lsun{L$_{\odot}$}

\def\arcs{\ifmmode {''}\else $''$\fi}
\def\arcm{\ifmmode {'}\else $'$\fi}
\def\parcs{\sa=.07em \sb=.03em
     \ifmmode $\rlap{.}$^{\scriptscriptstyle\prime\kern -\sb\prime}$\kern -\sa$
     \else \rlap{.}$^{\scriptscriptstyle\prime\kern -\sb\prime}$\kern -\sa\fi}
\def\parcm{\sa=.08em \sb=.03em
     \ifmmode $\rlap{.}\kern\sa$^{\scriptscriptstyle\prime}$\kern-\sb$
     \else \rlap{.}\kern\sa$^{\scriptscriptstyle\prime}$\kern-\sb\fi}

\def\Msun{M$_{\odot}$}

\def\spose#1{\hbox to 0pt{#1\hss}}
\def\simlt{\mathrel{\spose{\lower 3pt\hbox{$\mathchar"218$}}
     \raise 2.0pt\hbox{$\mathchar"13C$}}}
\def\simgt{\mathrel{\spose{\lower 3pt\hbox{$\mathchar"218$}}
     \raise 2.0pt\hbox{$\mathchar"13E$}}}
\def\lsim{\rlap{$<$}{\lower 1.0ex\hbox{$\sim$}}}
\def\gsim{\rlap{$>$}{\lower 1.0ex\hbox{$\sim$}}}

\begin{document}

\title{Observations of Ultraluminous Infrared Galaxies with the Infrared Spectrograph 
on the Spitzer Space Telescope II: The IRAS Bright Galaxy Sample\altaffilmark{1}}

\author{L. Armus\altaffilmark{2},
V. Charmandaris\altaffilmark{4,7},
J. Bernard-Salas\altaffilmark{3},
H.W.W. Spoon\altaffilmark{3},
J.A. Marshall\altaffilmark{3},
S.J.U. Higdon\altaffilmark{3},
V. Desai\altaffilmark{5},
H.I. Teplitz\altaffilmark{2},
L. Hao\altaffilmark{3},
D. Devost\altaffilmark{3},
B.R. Brandl\altaffilmark{6},
Y. Wu\altaffilmark{3},
G.C. Sloan\altaffilmark{3},
B.T. Soifer\altaffilmark{2,5},
J.R. Houck\altaffilmark{3},
T.L. Herter\altaffilmark{3}}

\altaffiltext{1}{based on observations obtained with the Spitzer Space Telescope, which is
operated by the Jet Propulsion Laboratory, California Institute of Technology, under NASA
contract 1407}
\altaffiltext{2}{Spitzer Science Center, MS 220-6, Caltech, Pasadena, CA 91125}
\altaffiltext{3}{Cornell University, Ithaca, NY 14853}
\altaffiltext{4}{University of Crete, Department of Physics, P.O. Box 2208 GR-71003, Heraklion, Greece}
\altaffiltext{5}{Division of Physics, Math \& Astronomy, California Institute of Technology, Pasadena, CA 91125}
\altaffiltext{6}{Leiden University, P.O. Box 9513, 2300 RA Leiden, The Netherlands}
\altaffiltext{7}{IESL/Foundation for Research and Technology - Hellas,
  GR-71110, Heraklion, Greece and Chercheur Associ\'e, Observatoire de
  Paris, F-75014, Paris, France}
\altaffiltext{8}{The IRS was a collaborative venture between Cornell University and
Ball Aerospace Corporation funded by NASA through the Jet Propulsion Laboratory and
the Ames Research Center}

\begin{abstract}

We present spectra taken with the Infrared Spectrograph$^{8}$ on Spitzer
covering the $5-38\mu$m region of the ten Ultraluminous Infrared Galaxies
(ULIRGs) found in the IRAS Bright Galaxy Sample.  Among the BGS ULIRGs, we
find a factor of 50 spread in the rest-frame $5.5-60\mu$m spectral slope.
The $9.7\mu$m silicate optical depths range from at least 
$\tau_{9.7} \le 0.4$ to $\tau_{9.7} \ge 4.2$,
implying line of sight extinctions of A$_{V} \sim 8$ mag to nearly 
A$_{V} \ge 78$ mag.  There is evidence for water ice and hydrocarbon
absorption and 
C$_{2}$H$_{2}$ and HCN absorption features in four and possibly six of the 10 BGS
ULIRGs, indicating shielded molecular clouds and a warm, dense ISM.  We have
detected [NeV] emission in three of the ten BGS
ULIRGs, at flux levels of $5-18\times 10^{-14}$erg cm$^{-2}$ sec$^{-1}$ and
[NeV] 14.3/[NeII] 12.8 line flux ratios of $0.12-0.85$.  The remaining BGS
ULIRGs have limits on their [NeV]/[NeII] line flux ratios which range from 
$\le 0.15$ to $\le 0.01$.
Among the BGS ULIRGs, the AGN fractions implied by either the
[NeV]/[NeII] or [OIV]/[NeII] line flux ratios (or their upper limits) are
significantly lower than implied by the MIR slope or strength of the $6.2\mu$m
PAH EQW feature.  Fitting the SEDs, we see evidence for hot (T $> 300$K) dust in
five of the BGS ULIRGs, with the 
fraction of hot dust to total dust luminosity ranging from $\sim1-23$\%, before 
correcting for extinction.
When integrated over the IRAC-8, IRS blue peakup, and MIPS-24 filter bandpasses,
the IRS spectra imply very blue colors for some ULIRGs at $z\sim1.3$, as the 
strong silicate absorption enters the MIPS-24 filter.  This is most extreme
for sources with significant amounts of warm dust and deep silicate absorption.
The large range in diagnostic parameters (line flux
ratios, spectral slopes, PAH strengths) among the nearest, brightest ULIRGs
suggests that matching deep survey results to a very small number of templates
may lead to biased results about the fraction of luminous, dusty starbursts and AGN at
high-z.

\end{abstract}


\section{Introduction}

Ultraluminous Infrared Galaxies (ULIRGs), i.e. those galaxies
with infrared luminosity L$_{IR} \gs 10^{12}$\lsun, have
the power output of quasars yet emit nearly all of their
energy in the mid and far-infrared part of the spectrum.
Most ULIRGs are found in interacting and merging systems
(e.g. Armus, Heckman \& Miley 1987; Sanders, et al. 1988a;
Murphy, et al. 1996), where the merger has driven
gas and dust towards the remnant nucleus, fueling a massive
starburst, and either creating or fueling a nascent AGN (Mihos
\& Hernquist 1996).  ULIRGs are rare in the local Universe,
comprising only $3\%$ of the IRAS Bright Galaxy Sample (Soifer
et al. 1987), yet at $z > 2-3$, ULIRGs may account for the bulk
of all star-formation activity and dominate the far-infrared background
(e.g. Blain et al. 2002).

Observations with the ISO satellite greatly expanded our
understanding of the mid-infrared spectra of ULIRGs (e.g.,
Genzel et al. 1998; Lutz et al. 1999; Rigopoulou et al. 1999;
Sturm et al. 2002; Tran et al. 2001).  Diagnostic diagrams
based on fine structure lines and aromatic emission features 
allowed some ULIRGs to be classified according to
their dominant ionization mechanism.  However, the complexities
of the ULIRG spectra, the fact that many are likely composite
AGN and starburst sources, and the limitations in sensitivity
of the ISO spectrometers, left many ULIRGs, even at relatively
low redshift, beyond the reach of these methods until now.

In order to adequately sample the local ULIRG population, and explore
the range in mid-infrared diagnostic tools for uncovering buried AGN,
we are obtaining mid-infrared spectra of a large number ($ > 100$) of
ULIRGs having redshifts of $0.02 < z < 0.93$ with the Infrared 
Spectrograph (IRS) 
on Spitzer, as part of the IRS team guaranteed
time program.  These sources are chosen primarily from 
the IRAS 1-Jy (Kim \&
Sanders 1998), 2-Jy (Strauss et al. 1992), and
the FIRST/IRAS radio-far-IR sample of Stanford et al. (2000).
In this paper, we present IRS spectra of the ten ULIRGs
in the IRAS Bright Galaxy Sample (Soifer et al. 1987).
The Bright Galaxy Sample (BGS) is a flux-limited, complete 
sample of all 324 galaxies with $60\mu$m IRAS flux densities 
greater than 5.4 Jy in the IRAS Point Source catalog (1985).
The ten ULIRGs in the BGS are IRAS 05189-2524, IRAS 08572+3915, 
IRAS 09320+6134 (UGC 5101),
IRAS 12112+0305, IRAS 12540+5708 (Mrk 231), IRAS 13428+5608 (Mrk 273), 
IRAS 14348-1447, IRAS 15250+3609, IRAS 15327+2340 (Arp 220), and 
IRAS 22491-1808.
An earlier reduction of the 
spectra of UGC 5101 was previously
presented in paper I of this series (Armus et al. 2004), and the silicate
absorption properties of 
IRAS 08572+3915 have been discussed in Spoon et al. (2006a).
The basic properties of the sample are listed in Table 1.
Throughout the paper, we will adopt a flat, $\Lambda$-dominated
Universe ($H_0 = 70$ km s$^{-1}$\ Mpc$^{-1}$, $\Omega_M=0.3$,
$\Omega_{\Lambda}=0.7$).

\section{Observations \& Data Reduction}

The ULIRGs were observed in the two low-resolution ($64 < R < 128$;
Short-Low and Long-Low or SL \& LL) and two high-resolution ($R\sim650$;
Short-High and Long-High or SH \& LH) IRS modules, using the Staring Mode
Astronomical Observing Template (AOT).  The galaxies were placed at the two
nod positions for each of the IRS slits.  High accuracy IRS blue peak-ups
were performed on nearby 2MASS stars for seven of the ten BGS sources,
before offsetting to the target galaxies.  For IRAS 12112+0305, and IRAS
14348-1447, and IRAS 22491-1808 we performed high-accuracy IRS blue
peak-ups on the nuclei themselves.  IRAS 08572+3915, IRAS 12112+0305, and
IRAS 14348-1447 have well-resolved double nuclei (see Condon et al. 1990,
Evans et al. 2000,2002).  In the case of IRAS 08572+3915, the IRS
observations were centered on the NW nucleus, which dominates the near and
mid-infrared emission (Soifer et al. 2000).  For IRAS 12112+0305 and IRAS
14348-1447, the slits were centered on the NE and SW nuclei, respectively.
Details of the observations are given in Table 2.  The IRS is fully
described in Houck et al. (2004).

All spectra were reduced using the S11 IRS pipeline at the Spitzer Science
Center.  This reduction includes ramp fitting, dark sky subtraction, droop
correction, linearity correction, and wavelength and flux calibration.
One-dimensional spectra were extracted from the IRS data products using the
SMART data reduction package (Higdon et al. 2004).  For the SH and LH data,
we performed full slit extractions.  For the SL and LL data, we used an
extraction aperture which expanded linearly with wavelength, but was set to
four pixels at the blue end of each order.  The SL and LL data have had
local background light subtracted by differencing the adjacent sub-slit
position (order), before spectral extraction.  The projected extraction
apertures in units of kpc are given in Table 1.  While the apertures
project to relatively large areas for the more distant systems, the slits
are always matched to the Spitzer beam and therefore, for compact MIR
sources such as ULIRGs, the flux we measure is dominated by the central
source.  Of the three sources with IRS peakup data, IRAS 22491-1808 is 
unresolved, while IRAS 12112+0305 and IRAS 14348-1447 are marginally resolved
at the $\sim10-15$\% level at $16\mu$m.  Mid-infrared images of IRAS 14348-1447 
in the ISOCAM 6.7 and $15\mu$m broadband filters which also show its resolved nature, have
previously been published by Charmandaris et al. (2002).
As a final step, we have normalized the SL and LL 1D spectra
upwards to match the IRAS $25\mu$m FSC data (Moshir et al. 1990).  Typical
scale factors were $1.1-1.2$ for the low-res data.  The SL and LL spectra
for all 10 sources are displayed in Fig.1.

Since the SH and LH slits are too small for on-slit background
subtraction, we have subtracted the expected background flux through
each slit based on the model of Reach et al. (2004).  The LH data
went through an additional cleaning to remove residual hot pixels,
which uses the B-mask (supplied with the BCD data) and the warm pixel
mask for the given IRS campaign.  Hot pixels were removed
by using the two nod positions and a measure of the 
light profile adjacent to the hot pixels, before extraction from 2D to
1D spectra.  For a full description of this method, see Devost et al.
(2006).  The SH and LH spectra were then scaled to the corresponding
IRS low-resolution spectra, using a single scale factor for each
module.  Typical scale factors were 1.2 for SH and 1.1 for LH.  The SH
and LH spectra are shown in Fig. 3 \& 4.  Noisy areas at the red end of the
SH and LH orders in order overlap regions are not shown or used in the
fitting process. In SH, these noisy areas amount to typically 10-30
pixels at the red end of orders 13-20, corresponding to an area of
decreased responsivity on the array.

\section{Results}

\subsection{Continuum \& Dust Features}

Broad, emission features associated with Poly-cyclic Aromatic Hydrocarbons
(PAHs) at 6.2, 7.7, 8.6, 11.3, and $12.7\mu$m, along with absorption from
amorphous silicates centered at 9.7 and $18\mu$m dominate the spectra in
nearly all cases (see Fig.1).  We also often detect the weaker PAH emission at 14.2,
16.4, $17.1\mu$m, and $17.4\mu$m.  The broad $17.1\mu$m feature sits under
both the $16.4\mu$m and $17.4\mu$m PAH features (see Smith et al. 2006) and 
the H$_{2}$ S(1) line which is often prominent in the ULIRG spectra (see below).
There is a large range in spectral shape
across the IRS wavelength range ($5-38\mu$m), among the sample galaxies.
IRAS 05189-2524 and Mrk 231 have relatively flat spectra with weak silicate
absorption and weak PAH emission.  At the other extreme are sources like
UGC 5101 and IRAS 14348-1447 which have very strong PAH emission, deep
silicate absorption, and water ice (most easily seen as a strong absorption
from $\sim5.5-6.5\mu$m, under the $6.2\mu$m PAH emission feature).  Water ice
absorption is also evident in Arp 220, Mrk 273, and IRAS 15250+3609.  This
absorption, suggested by Spoon  et al. (2002) to be indicative of shielded
molecular clouds along the line of sight to the nucleus, appears to be
common among nearby ULIRGs, appearing in half of the BGS sources.  
Note that
although the $5.5-6.5\mu$m absorption is dominated by water ice, hydrocarbon
absorption at $6.85\mu$m and $7.25\mu$m is often evident redward of the $6.2\mu$m 
PAH emission feature (see Spoon et al. 2006b).
IRAS 08572+3915 has a very interesting spectrum, since it shows extremely
strong silicate absorption at both 9.7 and $18\mu$m but no PAH emission.
For the BGS sample as a whole, the silicate optical depths range from
$\tau_{9.7} \le 0.4$ (A$_{V} \le 7.9$ mag) in  IRAS 05189-2524, to
$\tau_{9.7} = 4.2$ (A$_{V} =78$ mag) in IRAS 08572+3915, adopting a smooth
continuum anchored at $5.3-5.6\mu$m, $14\mu$m (for the PAH-dominated 
sources), and $34\mu$m, and
A$_{V}$/$\tau_{9.7} = 18.5$ from Roche \& Aitken (1984).  Optical 
depths measured in this way are always lower limits to the ``true" 
optical depth (see below).  The IRS
low-resolution spectra of the ten BGS ULIRG are shown together in Fig.  2,
normalized at $60\mu$m (Fig. 2a) and $25\mu$m (Fig. 2b).  In both
normalizations, IRAS 08572+3915 has the flattest (bluest), and Arp 220 has
the steepest (reddest) spectrum of of the BGS ULIRG sample.  There is
nearly a factor of 50 (25) spread among the ULIRGs at $5.5\mu$m when
normalized at $60\mu$m ($25\mu$m).

We also detect absorption features from gas-phase interstellar C$_{2}$H$_{2}$ and HCN
at $13.7\mu$m and $14.0\mu$m, respectively, in four of the ten BGS ULIRGs,
namely Mrk 231, Arp 220, IRAS 08572+3915, and IRAS 15250+3609.  Both
absorption features are detected in all galaxies, although the HCN line in
Mrk 231 is uncertain due to a continuum ``jump", just redward of the
absorption.  These features are labelled in Fig.3.  Weak C$_{2}$H$_{2}$
absorption may also be present in IRAS 05189-2524 and Mrk 273,
but in these galaxies the HCN line is not detected.
Fluxes and EQWs
for the C$_{2}$H$_{2}$ absorption line range from $5-10 \times 10^{-21}$W
cm$^{-2}$ s$^{-1}$ and $0.002-0.012\mu$m, respectively, with IRAS
15250+3609 having the deepest C$_{2}$H$_{2}$ absorption and Mrk 231 having
the shallowest absorption among the four ULIRGs.  Fluxes and EQWs for the
HCN line are smaller, ranging from $2-6 \times 10^{-21}$W cm$^{-2}$
s$^{-1}$ and $0.002-0.007\mu$m, respectively, again with IRAS 15250+3609
showing the deepest absorption.  The strengths of the  C$_{2}$H$_{2}$ and HCN
features, as well as the measured $\tau_{9.7}$ are listed for the BGS ULIRGs in 
Table 5.  The strengths of these features, when fit
to models which vary the excitation temperature and column density of the
gas, can be used to constrain the temperature of the gas and, in some
cases, its chemical evolutionary state (e.g.  Lahuis \& van Dishoeck 2000,
and Booonman et al. 2003). A detailed analysis of these lines in ULIRGs
will be presented in a future paper (Lahuis et al. 2006).  However, the positions of the line
centroids at $13.7\mu$m and $14.0\mu$m already suggest that a warm gas component
of $\sim200-400$K dominates the absorption in Mrk 231, Arp 220, IRAS
08572+3915 and IRAS 15250+3609.

\subsection{Emission Lines}

The high-resolution (SH and LH) ULIRG spectra are dominated by unresolved
atomic, fine-structure lines of Ne, O, Si, and S, covering a large range in
ionization potential (see Figs. 3 \& 4 and Tables 3 \& 4).  Ratios of the fine
structure lines can be used to gauge the dominant ionizing source - either
hot stars or an active nucleus.  Some features, e.g. the [NeV] lines at
14.3 and $24.3\mu$m, imply the presence of an AGN by their very detection
in a galaxy spectrum, since it takes 97.1 eV to ionize Ne$^{3+}$, and this
is too large to be produced by O stars.  [NeV] emission is seen in the
spectra of individual Galactic planetary nebulae (Bernard-Salas et al.
2001, Pottasch et al. 2001), but not in the integrated nuclear
(many kpc) spectrum of a galaxy unless an AGN is present.  The same is not
true for [OIV], since it takes only 55 eV to ionize O$^{2+}$.  The $25.9\mu$m
line has been seen in a number of pure starburst galaxies (e.g., Lutz et al.
1998, Smith et al. 2004, Devost et al. 2006).

Of the 10 BGS ULIRGs, we have detected [NeV] emission in three: IRAS
05189-2524, Mrk 273, and UGC 5101.  The detection of [NeV] in UGC 5101 has
been previously reported by Armus et al. (2004).  The fluxes in the
$14.32\mu$m lines are 18.36, 12.86, and $5.10\times 10^{-14}$ erg cm$^{-2}$
s$^{-1}$, in IRAS 05189, Mrk 273, and UGC 5101, respectively.  In all three
galaxies we detect both the $14.32\mu$m and $24.32\mu$m [NeV] lines.  
The $14.32\mu$m emission line was also seen in
the SWS spectrum of Mrk 273 (Genzel et al. 1998) at a level of $8\times
10^{-14}$ erg cm$^{-2}$ s$^{-1}$, about 50\% less than our measured flux.
For sources undetected in [NeV] with both the SWS (Genzel et al.) and the
IRS, the limits are typically a factor of $\sim 10-20$ lower in the IRS
data than the SWS data.  The nearby, low-ionization [ClII] feature at
$14.37\mu$m can easily be mistaken for [NeV], but the resolution of the IRS
and the high S/N of our data allow us to separate out the two lines in all
cases.  We have detected the [ClII] line in 5/10 of the BGS ULIRGs.  In
some cases (e.g. Mrk 273) both the [NeV] and the [ClII] lines are detected.
In others (e.g. Arp 220) only the [ClII] line is seen.

The [SIII] 18.71/[SIII] 33.48 line flux ratio can be used as a diagnostic
of the density of the ionized gas. For the BGS ULIRGs, this ratio ranges
from a lower limit of $\ge 0.17$ in Arp~220, to $\sim 0.65$ in IRAS
22491-1808, implying electron densities ranging from the low density limit
(for gas at T$= 10^{4}$ K) to about $200-300$ cm$^{-3}$. The average [SIII]
18.71/[SIII] 33.48 line flux ratio for the BGS ULIRGs with both lines
detected is 0.51, implying an average electron density of about $100-200$
cm$^{-3}$.  Our average ratio is consistent with that found for most of the
starburst galaxies analyzed by Verma et al. (2003).  It is important to
note, however, that the [SIII] line fluxes have not been corrected for
extinction, so the intrinsic [SIII] 18.71/[SIII] 33.48 line flux ratio and
the corresponding densities, are likely to be higher than reported here.
We do not know the extinction to the ionized gas, but an A$_{V} = 10$ mag
would cause us to understimate the [SIII] line flux ratio, and hence the
densities, by up to a factor of $\sim2-3$.

In addition to the fine-structure lines, the pure rotational lines of
H$_{2}$ (e.g., S(3) 9.66, S(2) 12.28, and S(1) 17.03) are often very strong
in the BGS ULIRGs.  The H$_{2}$ lines can be used to determine the
temperature and mass of (warm)
molecular gas.  All 10 BGS ULIRGs can be fit with at least one temperature
component, and three galaxies (Arp 220, IRAS 12112+0305, and IRAS 14348-1447)
require two components, due to the detection of the S(7) lines which indicate
the presence of a hot (T$\sim 1200$ K) component.  From fits to the S(1) through
S(3) lines, the BGS ULIRGs have molecular gas temperatures of about $260-360$K
and masses of about $3\times 10^{7}$\Msun (IRAS 08572+3915) up to 
$120\times 10^{7}$\Msun (IRAS 12112+0305).  This is typically a few percent 
of the cold molecular gas mass as derived from $^{12}$CO obsevations.
The hot ($\sim 1200$K) gas which dominates 
the S(7) emission typically has a mass which is a factor of $50-100$ lower than 
the gas at $\sim 300$K.  The state of the warm
molecular gas in our larger sample of ULIRGs, including the BGS ULIRGs,
is discussed fully in Higdon et al. (2006).

\subsection{Diagnostic Diagrams}

We present [NeV] 14.3/[NeII] 12.8 and [OIV] 25.9/[NeII] 12.8 diagnostic
emission line flux ratio diagrams for the ULIRGs in Fig. 5 \& 6.  In both
cases the fine structure line ratio is plotted against the $6.2\mu$m PAH
equivalent width (EQW) in microns, and a sample of AGN and starburst
galaxies observed with the IRS are shown for comparison (Devost et al.
20006, Brandl et al. 2005, Weedman et al. 2005, Hao et al. 2005).  The AGN
included are I Zw1, NGC 1275, Mrk 3, PG0804+761, PG1119+120, NGC 4151,
PG1211+143, 3C273, CenA, Mrk 279, PG1351+640, Mrk 841, and PG2130+099.  The
starbursts included are NGC 660, NGC 1222, IC342, NGC 1614, NGC 2146, NGC
3256, NGC 3310, NGC 4088, NGC 4385, NGC 4676, NGC 4818, NGC 7252, and NGC
7714.  The ULIRGs Mrk 1014, Mrk 463e (Armus et al. 2004) and NGC 6240
(Armus et al. 2006) have been added to the BGS ULIRGs in Fig. 5 \& 6.
Since both [NeV] and [OIV] are much stronger in AGN than starburst
galaxies, while the $6.2\mu$m PAH feature is weak or absent in AGN, these
diagnostic diagrams can effectively separate galaxies dominated by an AGN
from those dominated by a starburst (e.g.  Genzel et al. 1998, Sturm et al.
2002).  

In the simplest interpretation of Fig.5 \& 6, ULIRGs scatter between the
location of pure AGN (detected [NeV], little or no PAH emission) and pure
starbursts (strong PAH emission, no [NeV], and weak or absent [OIV]).  Hot
dust reduces the apparent strength of the $6.2\mu$m PAH EQW, even if
vigorous star formation is present.  In both the [NeV] and [OIV] diagrams,
we also indicate simple, linear mixing lines (i.e.,  Sturm et al. 2002)
along both axes which are meant to roughly indicate the percentage of AGN
and starburst contributing to the emission in a particular object.  The
vertical red lines (along the [NeV]/[NeII] or [OIV]/[NeII] axes) indicate
the percentage of the emission that is being generated by an AGN.  In both
cases the 100\% level is taken to be the average of the line ratio in the
AGN with detected emission.  As the fractional contribution from the AGN
decreases, (or equivalently the fractional contribution from a starburst,
increase), the [NeV]/[NeII] and [OIV]/[NeII] line flux ratios drop.  The
horizontal green lines (along the $6.2\mu$m axes) indicate the percentage
of the emission that is being generated by a starburst.  As the fractional
contribution from the starburst drops (or equivalently the fractional
contribution from an AGN increases), the PAH EQW gets smaller.  Here the
100\% level is taken to be the average of the $6.2\mu$m EQW in the pure
starburst galaxies.  The drop in $6.2\mu$m EQW with decreasing starburst
(increasing AGN) fraction is probably due to both a decrease in the PAH
emission, and an increase in the amount of hot dust continuum (see below).
There is significant scatter in the [NeV]/[NeII] and [OIV]/[NeII] line flux
ratios among the AGN themselves, some of which may be due to
circumnuclear star-formation (although they generally have very low $6.2\mu$m PAH
EQWs), and some of which is likely to be intrinsic.  Ionization models (Voit 1992)
suggest a large scatter in the line flux ratios 
due to variations in ionization parameter and
spectral index. 

There is a large range in [NeV]/[NeII] ($< 0.015$ to 1.58), [OIV]/[NeII]
($< 0.051$ to 6.23), and 6.2 PAH EQW (0.006 to $0.517\mu$m) among the 13
ULIRGs depicted in Fig. 5 \& 6.  In both plots, Mrk 463e (IRAS 12112+0305)
has the largest (smallest) line flux ratio, although strictly speaking IRAS
12112+0305 has only the most stringent upper limit, since it is undetected
in both [NeV] and [OIV].  IRAS 22491-1808 has the largest 6.2 PAH EQW,
$0.59\mu$m, among the sample ULIRGs.  In both the [NeV]/[NeII] and
[OIV]/[NeII] excitation diagrams, the ULIRGs scatter between the AGN and
the starburst galaxies.  The objects which would be considered pure AGN,
and for which the active nucleus dominates the bolometric luminosity,
generally have [NeV]/[NeII] line flux ratios of $1-3$, and $6.2\mu$m
equivalent widths less than $0.005-0.02\mu$m.  The starburst galaxies have
[NeV]/[NeII] ratios ($6.2\mu$m PAH EQWs) which are generally a factor of
$50-100$ smaller (larger) than the pure AGN values.  The linear mixing
lines suggest that the ULIRGs Mrk 463e, IRAS 05189-2524, and Mrk 1014 are
dominated by central AGN, since they have AGN contributions above 50\% in
[NeV]/[NeII], perhaps not surprising since all are optically classified as
Seyfert galaxies.  While IRAS 05189-2524 and Mrk 1014 have PAH emission, it
is at least an order of magnitude weaker than in the starburst galaxies,
and therefore young stars likely contribute less than 10\% of the power in
these ULIRGs.  The [OIV]/[NeII] ratio in IRAS 05189-2524 suggests an AGN
contribution a bit lower than does the [NeV]/[NeII] ratio ($\sim$30\% vs
60\%), but given the uncertainties and the extremely weak $6.2\mu$m PAH
emission, the AGN dominating the energetics is likely in this source.

Conversely, ULIRGs such as IRAS 12112+0305 and IRAS 22491-1808, which have
large $6.2\mu$m EQWs ($0.52\mu$m and $0.59\mu$m, respectively), comparable
to the starburst galaxies, and no [NeV] detections, are likely to have
their bolometric luminosities dominated by young stars.  The same is true
for both IRAS 14348-1447 and Arp 220, since both have very stringent limits
on [NeV], although their $6.2\mu$m PAH EQWs are about half that seen in the
majority of the pure starburst galaxies.  IRAS 12112+0305, IRAS 14348-1447 and Arp 220 are
all characterized as starbursts from their L-band spectra, since they have
relatively strong $3.3\mu$m PAH emission (Imanishi, Dudley \& Maloney 2006).
Both Mrk 273 and UGC 5101 have
[NeV], [OIV], and $6.2\mu$m PAH EQWs that fall in-between
pure starbursts and AGN, suggesting that a central engine and young stars
contribute measureably to their luminosity.  Both galaxies have relatively
weak $6.2\mu$m emission (after correcting for water ice absorption which 
is very strong in UGC 5101), suggesting that star formation contributes about
30\% of the luminosity.  However, their [NeV] and [OIV] line fluxes are also low
(especially the [OIV] line flux in UGC 5101), implying that the AGN
contribution is only about 20-40\% in Mrk 273 and less than 10\%
in UGC 5101.  Mrk 273 is listed as AGN-dominated in Imanishi et al. (2006)
based upon it's L-band spectrum.

Finally, there are three BGS ULIRGs, Mrk 231, IRAS 08572+3915, and IRAS 15250+3609
with seemingly inconsistent positions in Fig.5 \& 6.  These ULIRGs have
very weak $6.2\mu$m PAH EQWs - at least a factor of $40-50$ below that seen
in pure starburst galaxies, yet none have any detectable [NeV] or [OIV]
emission.  It is certainly true that the limits on [NeV] emission in these
galaxies are not nearly as stringent as other ULIRGs in our sample (e.g.
IRAS 12112+0305), but their [NeV]/[NeII] emission line flux ratios are
still at least a factor of $5-10$ below those seen in the pure AGN.  Mrk
231 is an optically classified broad-line AGN, whereas IRAS 08572+3915 and IRAS
15250+3609 are LINERs.  While IRAS 15250+3609 has no published
L-band spectra, both Mrk 231 and IRAS 08572+3915 are classified as AGN-dominated
based on extremely weak $3.3\mu$m emission (Imanishi, Dudley \& Maloney 2006).

The same general trend of the ULIRGs scattering between pure AGN and pure
starburst galaxies can be seen in Fig.7, where we plot the ratio of the
[NeV] $14.3\mu$m and IR flux, both uncorrected for reddening, against
the IRAS, far-infrared 25/60 micron flux density ratio.  Here, the $8-1000\mu$m IR flux
is derived from the IRAS data, using the prescription of Perault (1987).
The 25/60 micron ratio has often been used to separate ULIRGs
which are ``warm" ($25/60 > 0.2$) and possibly AGN-dominated, from those
that are ``cold" ($25/60 \le 0.2$) and likely to be dominated by star
formation (De Grijp et al. 1985; Sanders et al. 1988b).  In general, ULIRGs with warmer colors
have larger [NeV]/FIR, ranging from a low of less than about $0.02 \times
10^{-4}$ in Arp 220, to a high of nearly $5 \times 10^{-4}$ in Mrk 463e.
Since the extinction to the [NeV]-emitting region is unknown (see below),
part of the scatter in Fig.7 may be due to the fact that we have not
corrected the [NeV] flux.  However, the extremely large range in the
[NeV]/FIR ratios among AGN-dominated ULIRGs (Mrk 1014, Mrk 231, Mrk 463, IRAS 05189-2524)
and the AGN population itself,
suggests that extinction is not the only factor.  Notice also, that sources
with very weak [NeV] emission (e.g. NGC 6240 and UGC 5101) have larger
[NeV]/FIR ratios than Mrk 231, even though the lines of sight to the nuclei
in the former two ULIRGs are heavily obscured.

Another way to probe the presence of an active nucleus 
is to trace the presence of an AGN torus directly via the thermal emission
from hot dust. It is known both from theoretical models (e.g., Pier \&
Krolik 1992; Granato et al. 1997) and direct observations (e.g.,
Alsonso-Herrero et al. 2001), that the AGN radiation field can heat dust up
to temperatures of $\sim1000$K for silicate and $\sim1500$K for
graphite grains. As a result, the dust continuum emission becomes prominent
at short infrared wavelengths (3-6\,$\mu$m).

Such a diagnostic method was proposed by Laurent et al. (2000) based on
spectral data from the Infrared Space Observatory. Here, we apply this
approach to our Spitzer/IRS data making only minor modifications. As
discussed in detail in Laurent et al., the diagnostic assumes that the
integrated mid-IR emission can be represented by a sum of contributions
from (1) regions where the dust is predominantly heated by an AGN and (2)
regions where star formation is the main source of energy. Both AGN and
extreme starbursts contribute to the destruction of PAHs as well as to the
transient heating of the very small grains, leading to a strong continuum
at $14-15\mu$m. One can select three template spectra, one of a
dominant AGN source, one of a typical HII region and one of a PAH dominated
photo-dissociation region (PDR), and attempt to identify the relative
contribution of each template to any measured spectrum.  If we use the ratio of the
6.2$\mu$m PAH flux to the underlying continuum flux at 5.5$\mu$m as a tracer of
the contribution of the quiescent star formation and PDRs to the integrated
galaxy flux, then both AGN and HII regions will display a very low value.
Similarly, the ratio of the 5.5$\mu$m to the 15$\mu$m flux will be low for
starbursts, since they lack the extreme radiation field which heats up the
dust (presumably in a torus) to high temperatures.
In Fig. 8, we present a revised version of the Laurent et al. 
diagram using the 15$\mu$m to 5.5$\mu$m flux ratio (15/5.5) versus the 6.2PAH
over 5.5$\mu$m continuum ratio (6.2/5.5).  Here, 3C273 (Hao et al. 2005) is chosen to
represent the ``pure" AGN (although this object only has an upper limit on the 
measured $6.2\mu$m PAH flux), while M17 and NGC 7023 are chosen to define the
HII region and PDR vertices, respectively.  ISO SWS data 
are used for M17 and NGC 7023 (Peeters, Spoon \& Tielens 2004).  As in Figs. 5 and 6, AGN,
starbursts, and ULIRGs measured with the IRS are overplotted for comparison.
Simple linear mixing lines are indicated between each vertex on the plot.
For any position, one can then estimate the relative fraction of AGN and 
young stars to the dust heating.  Just as in the excitation diagrams based
upon the [NeV] and [OIV] emission lines, the ULIRGs scatter between pure
AGN and pure starbursts in Fig.8.  However, in this diagram, the ULIRGs
appear on average to have much larger AGN fractions, due to their low 
15/5.5 and 6.2/5.5 flux ratios.  Note however, that the pure starburst
galaxies also have low 15/5.5 flux ratios as compared to HII regions, and 
at least five ULIRGs have 15/5.5 flux ratios consistent with the range
defined by the starbursts, which themselves have 15/5.5 about 1/10 as large
as the HII regions.  Most of the separation between AGN and starburst
galaxies in Fig.8 occurs in the 6.2/5.5 axis, where Mrk 463, Mrk 231, 
IRAS 08572+3915, IRAS 05189-2524, IRAS 15250+3609, and Mrk 1014 are at
least 90\% AGN.

\subsection{SED Fitting}

To estimate the effect of dust emission at different temperatures on the
ULIRG SEDs, and to measure the strengths of the PAH emission features
against the underlying, silicate-absorbed continuum, we have fit the
spectra with a multi-component model which includes $2-3$ graphite and
silicate dust grain components, PAH emission features (fit with Drude
profiles), a 3500K blackbody stellar component, and unresolved Gaussian
emission lines (for the fine structure and H$_{2}$ lines).  The basic steps
in the fitting method are: (1) the creation of a PAH and continuum spectrum
by subtracting a first estimate of the emission lines from the observed
spectrum, (2) a first fit to the continuum and PAHs followed by a
subtraction of the continuum plus the estimated lines from the observed
spectrum, (3) a fit to the resulting PAH spectrum followed by a subtraction
of the PAH and continuum fit to create an observed emission line spectrum,
(4) a fit to the resulting spectrum to produce the next estimate of the
line strengths, and finally (5) a repeat of the process if the reduced
chi-sqauared is improved by more than 1\%.  The initial PAH template is
derived from the IRS spectrum of the starburst galaxy NGC 7714 (Brandl et
al. 2004).  We assume the PAH emission is unextinguished, yet the PAH
feature ratios are free to vary.  For a full description of the fitting
procedure, see Marshall et al. (2006).  To extend the SEDs, we have added
near-infrared and far-infrared data from Scoville et al. (2000), Moshir et
al.  (1990), Klaas et al. (2001) and Benford (1999) to the IRS
low-resolution spectra where available.  The results of the fitting are
presented in Table 6 and Figure 9.  While this sort of multi-component
fitting over such a large range in wavelength does not necessarily produce
a ``unique" fit to the data, the extremely high signal-to-noise IRS spectra
of the BGS ULIRGs do provide valuable constraints over the critical
$2-38\mu$m rest-frame range where warm dust is important.  In all cases we
determine the best fit by reducing the chi-sqaured residuals over the
entire spectral range.

In most cases the fit is extremely good.  In some cases, e.g.
IRAS 08572+3915 and IRAS 15250+3609, the detailed absorption profiles
(silicate, water ice, and hydrocarbon) are not matched.  For the water ice
and hydrocarbon absorptions which are evident between $5.5-6.5\mu$m, we have
used a template profile based upon the spectrum of UGC 5101 to fit all the
ULIRGs, and clearly there is sub-structure in the profiles among the ULIRGs
which is not accounted for by this method.  In the case of the silicate
absorption (dominating the continuum at $9.7\mu$m and $18\mu$m), some of
the profile mis-matches may be due to the presence of crystalline silicates.
The observed crystalline to amorphous
silicate fraction is known to vary among ULIRGs (Spoon et al. 2006a).  Our simple
model includes only amorphous silicates, and thus we expect some profile
mis-matches in the most heavily obscured sources.

For the BGS ULIRGs, and the four comparison sources listed in Table 6 (Mrk 1014,
Mrk 463e, NGC 6240 and NGC 7714 -- Armus et al. 2005, Marshall et al. 2006,
Brandl et al. 2004), we are asking a simple question: Does the IRS data, when
coupled with the NIR and FIR data from the literature, provide evidence for a
component of hot dust in the continuum, and if so, what is the fractional
contribution of this hot dust to the bolometric luminosity of the source ?  We
define a ``hot" dust component as one with a characteristic blackbody
temperature of T${_H} \ge 300$ K.  A spectral component with T$= 300$ K
represents grains with temperatures up to about $600$K, which is the temperature
expected for a graphite grain in thermal equilibrium at a distance of about 10pc
from an AGN source with L$=10^{12}$\lsun  -- near the upper
envelope usually estimated for the inner radius of a dusty torus surrounding an
AGN of this luminosity.  Although individual Galactic star-forming regions can produce 
very hot dust (e.g. van der Tak et al. 1999), the mid-infrared nuclear spectra of pure starburst
galaxies do not have measurable quantities of dust at this temperature
(Marshall et al. 2006, Brandl et al. 2005), so detection of hot dust in a nuclear
spectrum provides indirect evidence for a buried AGN.  Dust with T$_{H} \sim 700$K implies the
existence of significant numbers of grains at the sublimation temperature of graphite (about
1200K), which would emit at radii of about 1 pc from a central source of $10^{12}$\lsun.

We find evidence for a hot dust component in five out of the ten BGS
ULIRGs, namely Mrk 231, Arp 220, IRAS 05189-2524, IRAS 08572+3915, and UGC 5101. 
The best-fit temperature of the hot component ranges from
$\sim 450-850$K among these five sources.  The ratio of
the luminosity in the hot component to the total luminosity (uncorrected
for extinction) ranges from less than 1\%, in Arp 220, to 23\% in IRAS 08572+3915.
While the extinction to the hot dust component is fit in the modelling, the
correction is highly uncertain, ranging from a factor of about 1.5 in IRAS 05189-2524,
to a factor of five in UGC 5101.  Corrected values
of L$_{H}$/L$_{IR}$ are generally accurate to only $20-30\%$, and should
never exceed 1.0 (see column 7 of Table 6).  The extremely large correction
tabulated for IRAS 08572+3915 in Table 6 (more than a factor of ten) is probably an
over-estimate, brought on by a poor fit to the silicate absorption profile
(see Fig. 9) which may be (in part) caused by the presence of crystalline
silicates (Spoon et al. 2006a).  The remaining five BGS sources (Mrk 273,
IRAS 15250+3609, IRAS 22491-1808, IRAS 12112+0305, and IRAS 14348-1447) have no compelling
evidence for a measureable hot dust component.  While the addition of
another dust component does indeed lower the reduced chi-squared of the
fit, the temperature of this component is always below 250K, and in most
cases the ``hot" and ``warm" dust components overlap significantly when we
impose this third dust component on the fit.  In addition to the BGS sources, 
Mrk 1014, Mrk 463e, and NGC 6240 are also well-fit with a hot dust
component, and these are included in Table 6 as well as the fit to the 
starburst galaxy NGC 7714 (which has no hot dust emission).

As a way to measure the total ``hot" plus ``warm" dust luminosity in the
ULIRGs, which is less subject to the definition of precise temperature of the
components, we calculate the $1-40\mu$m dust luminosity as a function of the
total infrared luminosity for all sources in Table 6 (column 8).  The ratios range from a low
of 19\% in Arp 220 to a high of 81\% in Mrk 463e, before any corrections for
extinction.  This ratio generally measures the flatness of the spectrum,
taking care to exclude the prominent emission features (PAH and
fine-structure lines) and stellar emission in the NIR.  As expected, the
flattest BGS ULIRG spectra, namely Mrk 231, IRAS 05189-2524, and IRAS 08572+3915,
have the largest L$_{1-40}$/L$_{IR}$ luminosity ratios.  These
are also the sources with the largest hot dust luminosity ratios.  The
corrected values are generally very high ($35-100$\%), but again these
corrections are most uncertain in the galaxies with 
very deep $9.7\mu$m absorption features whose profiles are poorly fit.

Finally, we tabulate the PAH to total luminosity ratio for the BGS ULIRGs
and comparison sources in column 9 of Table 6.  These ratios are uncorrected for
extinction.  The largest ratio (4.4\%) is found in
NGC 7714, while the smallest is seen in Arp 220 and Mrk 463e (0.4\%).  The extremely
small L$_{PAH}$/L$_{IR}$ ratio for Arp 220 has been noted before (e.g.
Spoon et al. 2004).  Although Arp 220 is usually referred to as the ``prototypical"
ULIRG, the relative PAH strength is actually quite low among ULIRGs as a class,
especially those optically classified as starbursts.

\section{Discussion}

\subsection{AGN and Starburst Diagnostics}

The greatly increased sensitivity of the IRS and Spitzer, coupled with the
narrower slits of the IRS, provide better isolation of the nucleus against
the galactic disk than was possible with the ISO spectrographs, increasing
our ability to detect the signatures of weak (or obscured) AGN in the
mid-infrared.  Our regularly achieved flux limits after very short
integration times, and the detection of previously unseen AGN signatures in
some ULIRGs and starburst galaxies (Armus et al. 2004, 2006; Devost et al.
2006), attest to this advancement.  However, among the brightest ULIRGs
(the $60\mu$m selected BGS sample), we have detected high-ionization lines
in only three out of ten galaxies, namely UGC 5101, Mrk 273, and IRAS
05189-2524.  Of these three, one (Mrk 273) had a previous [NeV] detection
with the SWS (Genzel et al. 1998), and two are optically-classified as
Seyfert 2 galaxies (Mrk 273 and IRAS 05189-2524).  Fitting the SEDs, we see
evidence for hot (T $> 300$K) dust in 5/10 of the BGS ULIRGs, namely UGC
5101, IRAS 05189-2524, Mrk 231, IRAS 08572+3915, and Arp 220.  The
temperature of this hot dust is highest in UGC 5101 and Arp 220, but also
the fraction of the total luminosity is low in these sources, especially in
Arp 220 where it is $\le 1\%$.  Although Mrk 273 has high-ionization
emission lines, the spectral shape is relatively steep, and the addition of
a hot dust component does not significantly improve the fit.  If we add the
three nearby, bright ULIRGs from Armus et al. (2004, 2005), Mrk 1014, Mrk
463e and NGC 6240, all of which have high-ionization lines and hot dust, we
have 6/13 with [NeV] emission and 8/13 with hot dust.

In Table 7 we have assembled classifications for the optical/near-infrared,
mid-infrared (IRS), and X-ray for the 10 BGS ULIRGs along Mrk 1014, Mrk
463e, and NGC 6240.  The IRS classifications are based upon the presence
and strength of [NeV] and [OIV] emission, the MIR spectral shape, and the
$6.2\mu$m and total PAH emission.  In general the [NeV] and [OIV] based AGN
fractions agree within a factor of two.  In IRAS 05189-2524 the [NeV] line
seems unusually strong (with respect to [OIV]) and in Mrk 273 and NGC 6240
the opposite is true ([NeV] appears weaker).  In most cases the AGN
fraction estimated from the high-ionization lines is much lower than that
estimated from the strength of the PAH emission or MIR slope.  While some
ULIRGs have very small AGN fractions as evidenced by a lack of [NeV], weak
[OIV], and strong PAH emission (e.g. IRAS 22491-1808), many sources,
including those with obvious AGN signatures in the optical or near-infrared
(e.g. Mrk 231, Mrk 273, and Mrk 1014) have weaker [NeV] and/or [OIV]
emission than expected given their measured PAH strength.  This is also
true for some of the ULIRGs classified as LINERs (e.g. UGC 5101 and NGC
6240) for which we have detected very weak [NeV] emission (from a buried
AGN) and PAH emission which is weaker than in pure starburst galaxies.  The
high-ionization tracers never suggest a larger AGN fraction than is
indicated by the strength of the PAH emission or the MIR slope.  
indicate that the coronal line region, as well as 
[OIV] emission that is powered by the AGN, either intercepts 
fraction of the available ionizing photons (either because of a low
extinction 
This dust must 
producing most of the [NeII] 
starburst.

Like Mrk 231 and IRAS 08572+3915, IRAS 15250+3609 has very weak PAH emission yet
no detectable [NeV].  In IRAS 15250+3609 the presence of water ice and
hydrocarbon absorption makes an accurate estimate of the $6.2\mu$m PAH emission
very difficult, and there is a possibility that we have understimated the
strength of this feature.  Like an extreme example of Arp 220, most of the
activity in IRAS 15250+3609 is probably deeply buried, even when observed in the
mid-infrared.  It is not surprising that this system also has the strongest
C$_{2}$H$_{2}$ and HCN absorption bands of the BGS sample.

An important caveat to using the [NeV]/[NeII] line flux ratio to
quantitatively estimate the AGN fraction in any particular source, is that
the line ratio is not independent of extinction, even though the [NeV] and
[NeII] lines are very close in wavelength.  The [NeV] emission comes from a
region of hot, tenuous gas surrounding the central source, known as the
Coronal Line Region (CLR).  The [NeII] on the other hand, can have a
significant contribution from an extended, foreground starburst.  This may
partly explain the extremely low [NeV]/[NeII] levels in UGC 5101 and NGC
6240.  In both of these ULIRGs the PAH emission suggests a much larger AGN
fraction than is apparent from the [NeV] emission.  Even correcting the
observed [NeV] emission for the extinction implied by the silicate optical
depths, $\tau \sim 1.2$ (or A$_{V} \ge 22$ mag) for NGC 6240 and $\tau \sim
1.6$ (or A$_{V} \ge 31$ mag) for UGC 5101, while substantial, moves neither
galaxy into the pure AGN area of Fig.5 or increases the [NeV] enough to
make the AGN fractions comparable to those estimated from the PAH emission
-- and this is assuming that all the correction is made to [NeV] and that
none of the [NeII] is obscured, which is highly unlikely.  So, while
extinction may be a factor in the measured [NeV]/[NeII] flux ratios and the
true AGN fractions may be higher than those listed in Table 7, for the
ULIRGs with extremely weak yet detected [NeV] emission (like UGC 5101 and
NGC 6240), the apparent silicate optical depths are not large enough to
derive corrected ratios which imply a significant AGN contribution to the
total power budget.

Of course, the apparent silicate optical depths, derived from smooth fits
to the continuum are always lower limits to the extinction, since they do
not take into account foreground continuum, PAH, or, more importantly,
silicate emission.  As discussed in Armus et al. (2006) in the context of
NGC 6240, the $\tau_{W}$ list in Table 6, which can be very large, may be a
more accurate estimate of the optical depth at $10\mu$m, since it does
explicitly correct for silicate emission.  For UGC 5101 and NGC 6240, using
the $\tau_{W}$ estimates to correct the measured [NeV] emission effectively
increases the estimated AGN fraction to $\sim80$\% and $\sim20$\%,
respectively, making both much more consistent with the estimates derived
from the PAH emission and MIR slopes.

Hard X-ray observations of NGC 6240 and UGC 5101 suggest a significant AGN
contribution, after correction for extremely large HI columns of $1-2\times
10^{24}$ cm$^{-2}$ (Vignati et al. 1999, Imanishi et al. 2003).  In fact,
NGC 6240 has the largest L$_{HX}$/L$_{IR}$ ratio of any ULIRG observed to
date, $\sim 250-740\times 10^{-4}$, comparable to those found for Seyfert
galaxies and Quasars (Ptak et al. 2003), suggesting that the central AGN
contributes as much as $50-100\%$ of the bolometric luminosity (see also
Lutz et al. 2003).  The large uncertainty in the extinction-corrected hard
X-ray flux is due to an uncertain correction for reflected light.  Seyfert
2 galaxies generally have values of $50-500\times 10^{-4}$, while Seyfert 1
galaxies have values of $200-2000\times 10^{-4}$ (Ptak et al. 2003).  The
third column of Table 7 gives the hard X-ray to infrared luminosity ratios
for the 13 ULIRGs discussed here.  Although there is a large spread, in all
cases the ULIRGs with significant AGN contributions (as measured either in
the optical/NIR or MIR spectra) have L$_{HX}$/L$_{IR}$ $ > 10-20\times
10^{-4}$.  The IRS starburst-like ULIRGs, including some with LINER-like
optical spectra, all have L$_{HX}$/L$_{IR}$ $< 1\times 10^{-4}$.  The ULIRG
with the largest [NeV]/[NeII] line flux ratio, Mrk 463e, has a relatively
low  L$_{HX}$/L$_{IR}$ ratio, for AGN.  While the CLR is relatively
unobscured in this Seyfert 2 ULIRG, apparently the accretion disk is
visible only behind significant HI - in this case, a column of about
$2\times 10^{23}$ cm$^{-2}$ (Bassani et al. 1999).

While there are sources with both high-ionization lines and hot dust, the
groups do not completely overlap, even though these are both tracers of a
(buried) AGN.  This can be explained by realizing that the [NeV] and the
hot dust are being generated in two different regions.  The hot dust comes
from either a shell at a radius of $1-10$ pc, or the inner edge of an
obscuring torus at a similar radius.  If the dust is in a torus which is
nearly edge-on, the coronal-line region, which gives rise to the [NeV]
emission, may actually be easier to observe in the mid-infrared if it has a
scale height which is larger than the thickness of the torus.  Evidence for
relatively large CLRs (tens of parsecs) in some nearby Seyfert galaxies,
has indeed been found (e.g., Maiolino et al. 2000, Prieto et al. 2005).  On
the other hand, if the torus is slightly tilted, affording us a view of the
hot, far side inner edge, both the [NeV] and the hot dust would be
observable.  However, an underlying uncertainty in the detectability of the
CLR is that both the extinction towards, and the covering factor of the
[NeV]-emitting clouds (i.e. what fraction of the ionizing radiation the
clouds intercept) is neither known nor constant from source to source.

Three of the ULIRGs (Mrk 231, Arp 220, and IRAS 08572+3915) have evidence
for hot dust, but no detected [NeV].  One source, Mrk 231, is anomalous. It
has an optically-detected broad-line region, a flat mid-infrared spectrum
(indicating hot dust), and yet no detected [NeV] emission.  While the broad
lines are polarized, and probably contain a significant scattered component
(Hines et al.  2001), it is unlikely that the absence of [NeV] can be
explained by extinction towards the CLR, since hot dust is observed, as
evidence by the shape of the mid-infrared spectrum, and this dust is likely
to be only a parsec or so from the central source.  In Mrk 231, the
covering factor of the nucleus by the coronal-line clouds may be very small
resulting in an unusually small or weak coronal-line region (compared to
other dusty AGN like Mrk 1014 or Mrk 463).  Alternatively the hot dust may
be coming from a warped disk just outside of the coronal-line region,
effectively blocking the [NeV] emission from our line of sight.  The second
source, Arp 220, is less surprising.  Spoon et al. (2004) suggest that much
of the power source remains hidden along our line of sight, even in the
mid-infrared.  We have fit the SED with a hot dust component, but the
contribution of this component to the total infrared luminosity is very
small, and therefore this detection remains uncertain.  We can obtain a
poorer fit to the data by removing the hot dust, and allowing the stellar
emission to be obscured by an A$_{V}$ $\sim 12$ mag, but since the hot dust
contribution is so small, either fit is still viable.  The third source
with evidence for hot dust, but no detectable [NeV] emission is IRAS
08572+3915.  This has perhaps the strangest mid-infrared spectrum of the
local ULIRGs. IRAS 08572+3915 has the largest silicate absorption and yet
the flattest mid-infrared spectrum (when normalized at either rest-frame
$60\mu$m or $25\mu$m) of the BGS ULIRGs.  The silicate profile suggests
crystalline silicates (Spoon et al.  2006a), with an A$_{V}$ of at least 78
mag.  Clearly there is an enormous amount of extinction towards the
nucleus, and yet a large amount of hot dust emission escapes without being
``downgraded" into the far-infrared.  L-band spectroscopy of this source
also shows strong absorption by dust, with little or no $3.3\mu$m PAH
emission (Imanishi, Dudley \& Maloney 2006).
Ground-based $10\mu$m and $20\mu$m spectrophotometry first led Dudley \&
Wynn-Williams (1997) to suggest that IRAS 08572+3915 harbored a buried AGN.
ULIRGs like IRAS 08572+3915 may be rare, representing a population of
buried AGN, not readily detectable by means other than their unusual
mid-infrared colors (see below).

\subsection {Mid-Infrared Colors}

Local galaxies display a wide range of mid-infrared spectral shapes due to the
presence of dust in emission and absorption.  In particular, the PAH emission
features at 6.2, 7.7, 8.6, 11.3, and $12.7\mu$m, and the silicate absorption
bands at 9.7 and $18\mu$m dominate the spectra of dusty galaxies over a large
range in luminosity.  Since these features are strong, and very broad, they can
be detected in broad-band photometry and serve as crude redshift indicators for
sources that are highly obscured in the optical, and even too faint for
mid-infrared spectroscopy with the IRS.  Recent studies of mid-infrared selected
galaxies have shown that low-resolution IRS spectra can be used to obtain
redshifts of $z=2-3$ ULIRGs with $24\mu$m flux densities, f$_{24} \ge 0.7$ mJy
and optical magnitudes of R$\ge 24-25$ mag (Houck et al.  2005, Yan et al. 2005).
Since the silicate absorption at $9.7\mu$m can often be the strongest feature in
the mid-infrared spectrum of a highly obscured galaxy, such as a ULIRG, a
combination of IRS blue peakup filter $16\mu$m imaging together with IRAC
$8\mu$m and MIPS $24\mu$m data can easily pinpoint the most obscured systems.
In particular, the 16/24 and 8/24 micron flux density ratios, when compared to
local templates, can be used to select obscured galaxies at $z = 1-2$.
Highly-obscured systems will exhibit a sharp increase in the 8/24 and 16/24 flux density
ratios as the $9.7\mu$m silicate absorption passes through the MIPS $24\mu$m
filter bandpass (Takagi \& Pearson 2005).  

As a way to ``calibrate" searches for $z\sim 1.5$ ULIRGs, we present the
8/24 and 16/24 micron color ratios for the BGS ULIRGs in Figs. 10 \& 11.
We also include the nearby starburst galaxy NGC 7714 (Brandl et al. 2004)
for comparison.  In all cases, the synthetic color ratio has been
calculated using the observed IRS spectra, the published IRS blue peakup
and MIPS $24\mu$m filter functions and, for the case of the 8/24 plot, the
ground-based NIR photometry used in the SED fitting.  The latter is
required because the blue edge of the IRAC $8\mu$m filter bandpass moves
below the short-wavelength cutoff of the IRS short-low second order
spectrum for $z > 0.3$.  The ULIRGs in both plots show qualitatively
similar behavior, namely a peak in the flux ratios at $z\sim0.3$ and
$z\sim1.3$ caused by the silicate absorption bands entering the MIPS
$24\mu$m filter.  However, the extremely blue colors predicted for some
ULIRGs at $z\sim1.3$ in the 16/24 flux ratio plot are striking.  The galaxy
with the strongest signal (bluest colors) in Fig. 11, IRAS 08572+3915, has
the deep silicate absorption at $9.7\mu$m and a relatively flat (warm)
mid-infrared SED.  Starburst-like ULIRGs, even those with strong silicate
absorption, typically have 16/24 colors of $1-2$ at $z =1.3$.  AGN-like
ULIRGs with little absorption are even redder.  IRAS 08572+3915, and to a
lesser extent, IRAS 15250+3609, have extreme colors, well above $16/24 =3$.
Galaxies such as IRAS 08572+3915, while rare, may be found in dedicated IRS
blue peakup and MIPS $24\mu$m imaging surveys.  Both Takagi \& Pearson
(2005) and Kasliwal et al. (2005) estimate that there may be up to a few
hundred so-called ``silicate drop-outs", per square degree, based upon
model predictions and the results of early Spitzer surveys.
It is the combination of a relatively flat, mid-infrared spectrum (indicating
significant amounts of hot dust) and deep silicate absorption that give
IRAS 08572+3915 its extreme mid-infrared colors.

\section{Summary}

In this paper we have presented mid-infrared $5-40\mu$m spectra of the ULIRGs in the
IRAS Bright Galaxy Sample, taken with the IRS on Spitzer.  Among this small sample 
there is a very large range in mid-infrared spectral properties (slope, absorption 
lines, emission lines) reflective of a variety of excitation conditions and ISM densities.
We find:

\noindent
(1) A factor of 50 spread in the spectral slope from $60\mu$m to $5.5\mu$m in the rest 
frame.  IRAS 08572+3915 is the flattest (bluest) while Arp 220 is the steepest (reddest)
among the BGS ULIRGs.

\noindent
(2) The $9.7\mu$m silicate optical depths range from $\tau_{9.7} \le 0.4$ in IRAS 05189-2524
to $\tau_{9.7} \ge 4.2$ in IRAS 08572+3915, implying line of sight (apparent) extinctions of 
A$_{V} \sim 7.9$ mag to more than A$_{V} \ge 78$ mag.  IRAS 08572+3915 has
both the flattest spectral slope and the deepest silicate absorption of the BGS ULIRG sample.

\noindent
(3) Evidence for water ice and hydrocarbon absorption (from $5.5-6.5\mu$m) in 7/10 BGS ULIRGs, 
as well as absorption features of C$_{2}$H$_{2}$ and HCN in four and possibly six of the
10 BGS ULIRGs, indicating shielded molecular clouds and a warm, dense ISM in these objects.

\noindent
(4) The presence of [NeV] emission at $14.3\mu$m and $24.3\mu$m in three out of the 10
BGS ULIRGs: IRAS 05189-2524, Mrk 273, and UGC 5101, with $14.3\mu$m line fluxes of 
18.36, 12.86, and $5.10 \times 10^{-14}$ erg cm$^{-2}$ s$^{-1}$, and [NeV] 14.3/[NeII 12.8 
line flux ratios of 0.85, 0.24, and 0.12, respectively.  The remaining seven BGS ULIRGs
have upper limits on their [NeV]/[NeII] line flux ratios which range from $\sim 0.01-0.15$.

\noindent
(5)  Fitting the SEDs, we see evidence for hot (T $> 300$K) dust
in 5/10 of the BGS ULIRGs, namely UGC 5101, IRAS 05189-2524, Mrk 231, Arp 220 and IRAS
08572+3915.  The fraction of hot dust to total dust luminosity is 
highest in IRAS 08572+3915 ($\sim23\%$, before correcting for considerable extinction), 
and lowest in Arp 220 ($\le 1\%$).  The temperature of this hot dust component
ranges from $\sim450-850$K.
Mrk 463e, Mrk 1014 and NGC 6240 also 
show evidence for hot dust in their spectra.

\noindent
(6) Constructing excitation diagrams from the [NeV]/[NeII], [OIV]/[NeII]
and $6.2\mu$m PAH equivalent width (EQW) suggests that Mrk 463e, Mrk 1014, 
and IRAS 05189-2524 are AGN dominated, while IRAS 12112+0305, IRAS 14348-1447, 
IRAS 22491-1808, and Arp 220 are starburst-dominated.  Mrk 273 appears to have
a significant starburst contribution ($30-50\%$) to it's total luminosity.
Mrk 231, IRAS 08572+3915, and IRAS 15250+3609 show very weak $6.2\mu$m PAH 
emission, and yet are undetected in [NeV] or [OIV], even though they are 
optically classified as a Seyfert 1 (Mrk 231), and LINERs (IRAS 08572+3915 
and IRAS 15250+3609), respectively.  Since both Mrk 231 and IRAS 08572+3915
have significant hot dust emission, it is likely that the coronal-line regions
in these sources are unusually weak or absent.  In the case of IRAS 15250+3609, 
the power source remains buried, even in the mid-IR.  Two sources, UGC 5101
and NGC 6240 have very weak [NeV] and hot dust emission, indicative of buried
AGN.  This is supported by the fact that
both of these sources have significant hard x-ray emission
behind extremely large columns of HI ($1-2\times 10^{24}$ cm$^{-2}$).
In all ULIRGs in this study, except for Mrk 463 and Mrk 1014, the AGN fractions implied by either
the [NeV]/[NeII] or [OIV]/[NeII] line flux ratios (or their upper limits) are significantly lower
than implied by the MIR slope or strength of the $6.2\mu$m PAH EQW feature.

\noindent
(7) Flux ratios between the IRS blue peakup and MIPS $24\mu$m filters, constructed
from the IRS spectra as a function of redshift for the BGS ULIRGs, imply significantly 
blue colors (large 16/24 flux ratios) at $z=1-2$, peaking at $z=1.3$, as previously 
suggested by Takagi \& Pearson (2005).  The one BGS ULIRG not observed with the ISO 
satellite, IRAS 08572+3915, has the most extreme colors
of any local ULIRG ($z \le 0.1$), due to a combination of extremely deep silicate absorption and
a flat SED in the mid-infrared.  Extreme 16/24 colors are also predicted for F00183-7111
(Spoon et al. 2004, Takagi \& Pearson 2005).  Although the BGS sample of ULIRGs is 
small, this suggests that perhaps up to $10\%$ of ULIRGs may have unusual colors indicative
deeply buried AGN.  These sources should be found from their IRS blue and MIPS $24\mu$m colors
in Spitzer imaging surveys now planned or underway.

Even among the nearest, brightest ULIRGs there is a startlingly large range 
in PAH equivalent width, high-ionization emission line flux ratios, and 
spectral slope.  We find evidence for AGN in the mid-infrared spectra 
of ULIRGs with optical/NIR Seyfert or LINER classifications.  We do 
not find evidence for buried AGN in ULIRGs
classified optically as starburst-like among the BGS sample.
However, the AGN fractions implied by the [NeV]/[NeII] or [OIV]/[NeII] line
flux ratios are often significantly lower than suggested by either the mid-infrared
slope or the strength of the $6.2\mu$m PAH feature.
Although we have yet to analyze the full IRS ULIRG sample, 
the idea of a ``typical" ULIRG mid-infrared spectrum appears much 
less useful a concept in light of the results of this survey thus far.

\acknowledgements

We would like to thank Bruce Draine, Aaron Evans, Dean Hines, David Hollenbach, 
Mark Lacy, Els Peeters, and Jason Surace
for many helpful discussions. 
Support for this work
was provided by NASA through an award issued by JPL/Caltech.

\begin{figure*}[lowres1]
\plotone{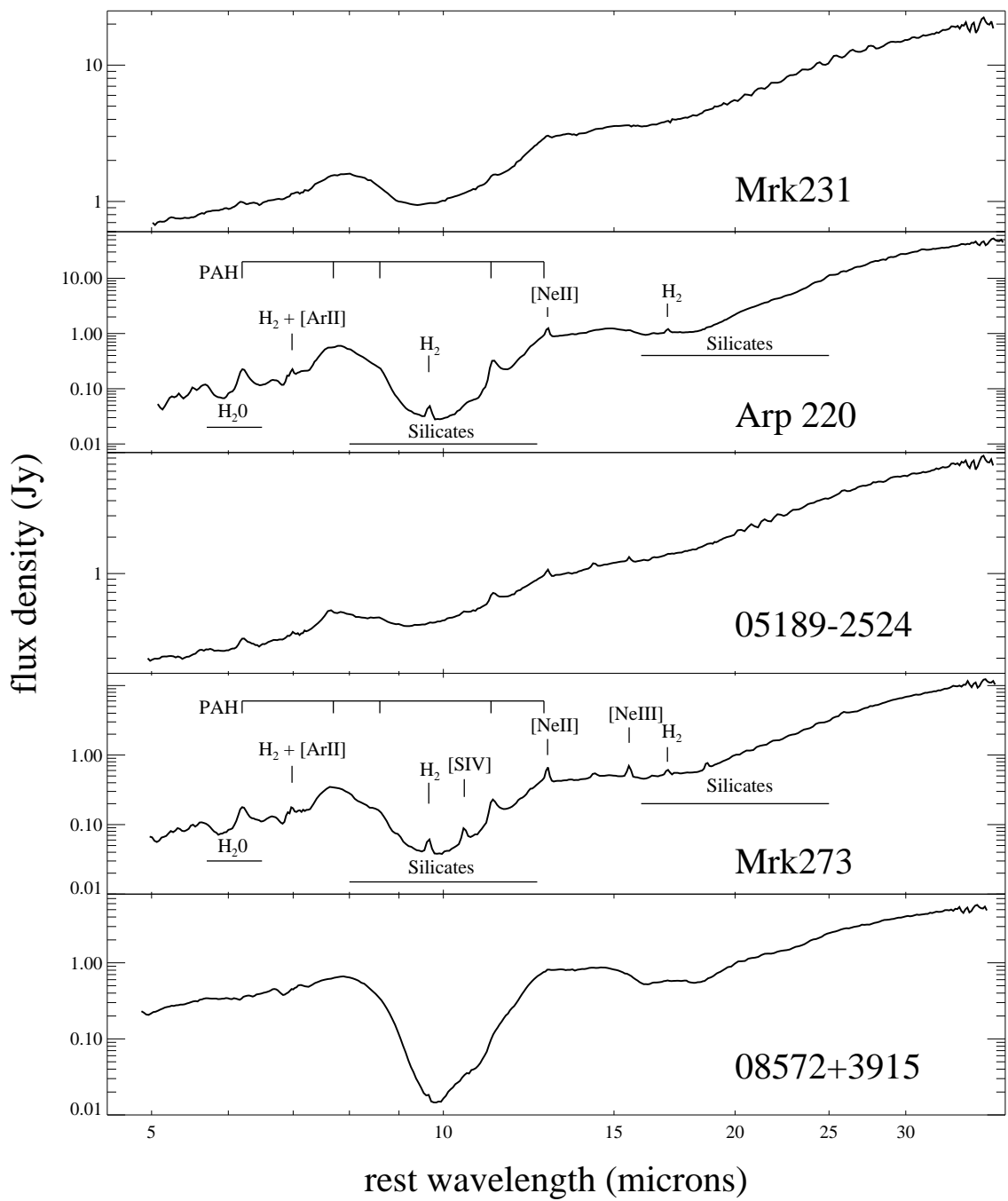}
\caption{IRS Short-Low \& Long-Low spectra of the 10 BGS ULIRGs in order
of decreasing $25\mu$m flux density.  Prominent 
emission features and absorption bands (the latter indicated by horizontal bars) are 
marked on representative spectra.  Not all features are marked on all spectra - see Tables 3 \& 4 for measured 
features.  Expanded views of the $5-14\mu$m regions (Short-Low) of each 
spectrum are shown at the end.}
\end{figure*}

\begin{figure*}[lowres2]
\plotone{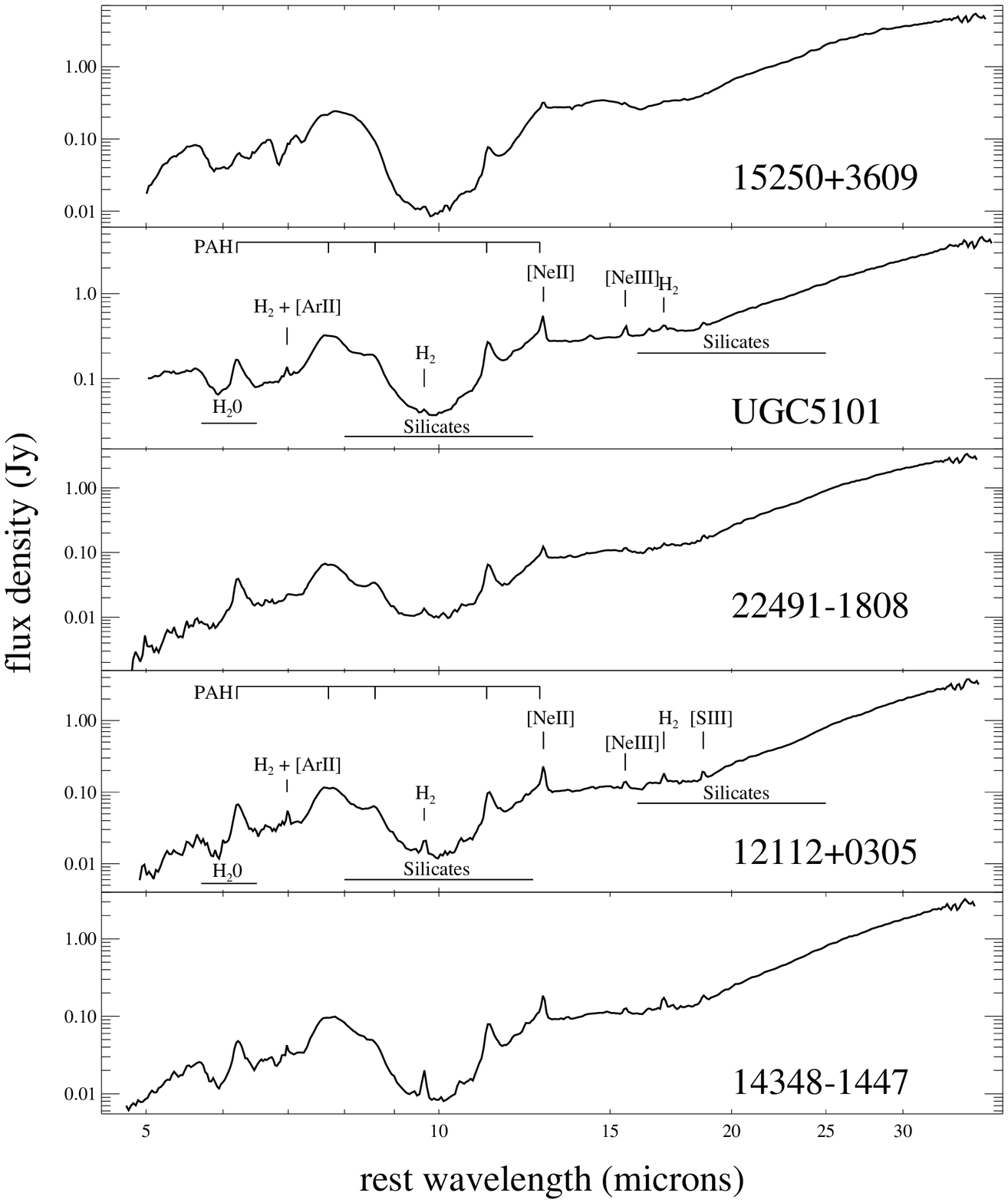}
\end{figure*}

\begin{figure*}[lowres1b]
\plotone{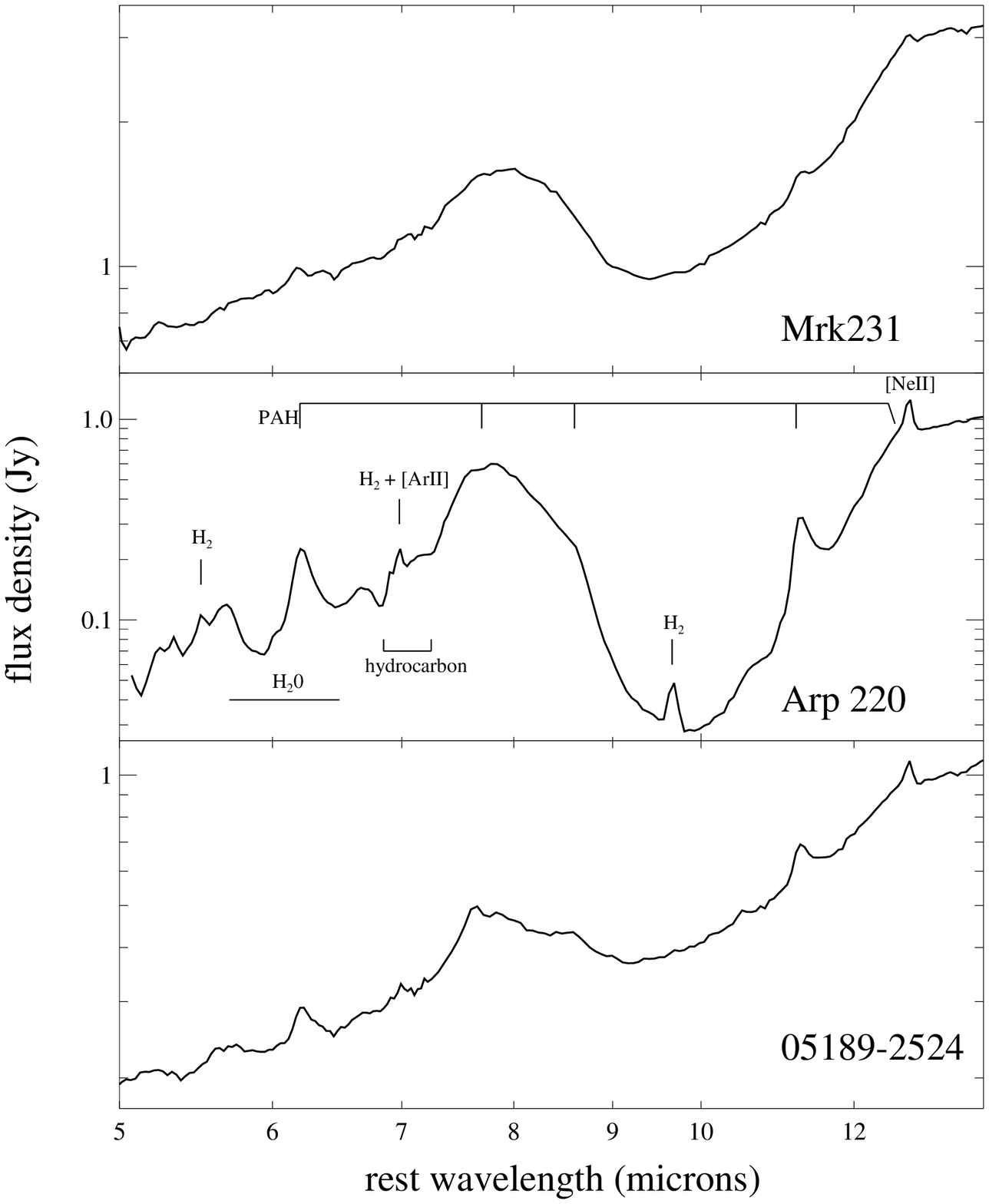}
\end{figure*}

\begin{figure*}[lowres1c]
\plotone{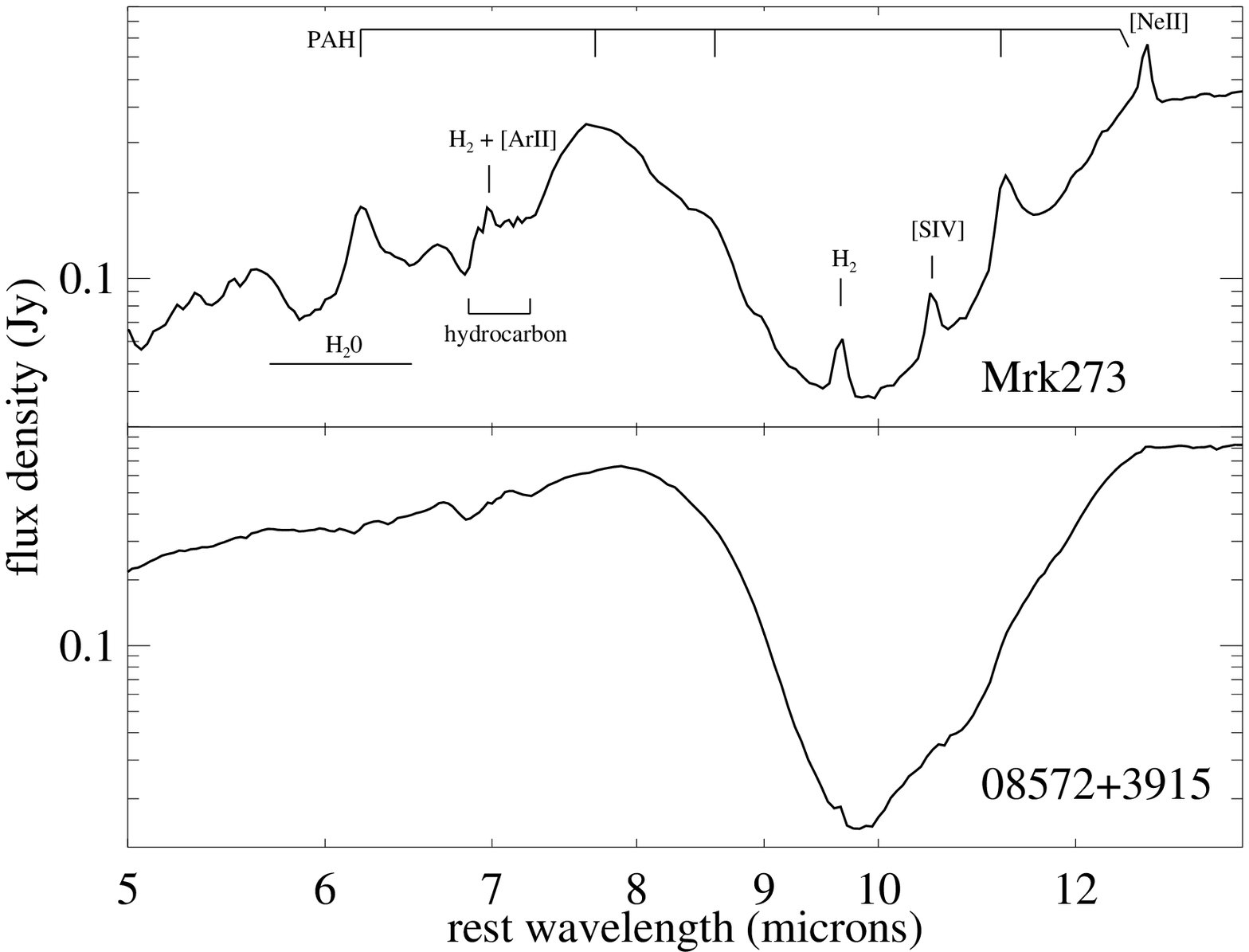}
\end{figure*}

\begin{figure*}[lowres2b]
\plotone{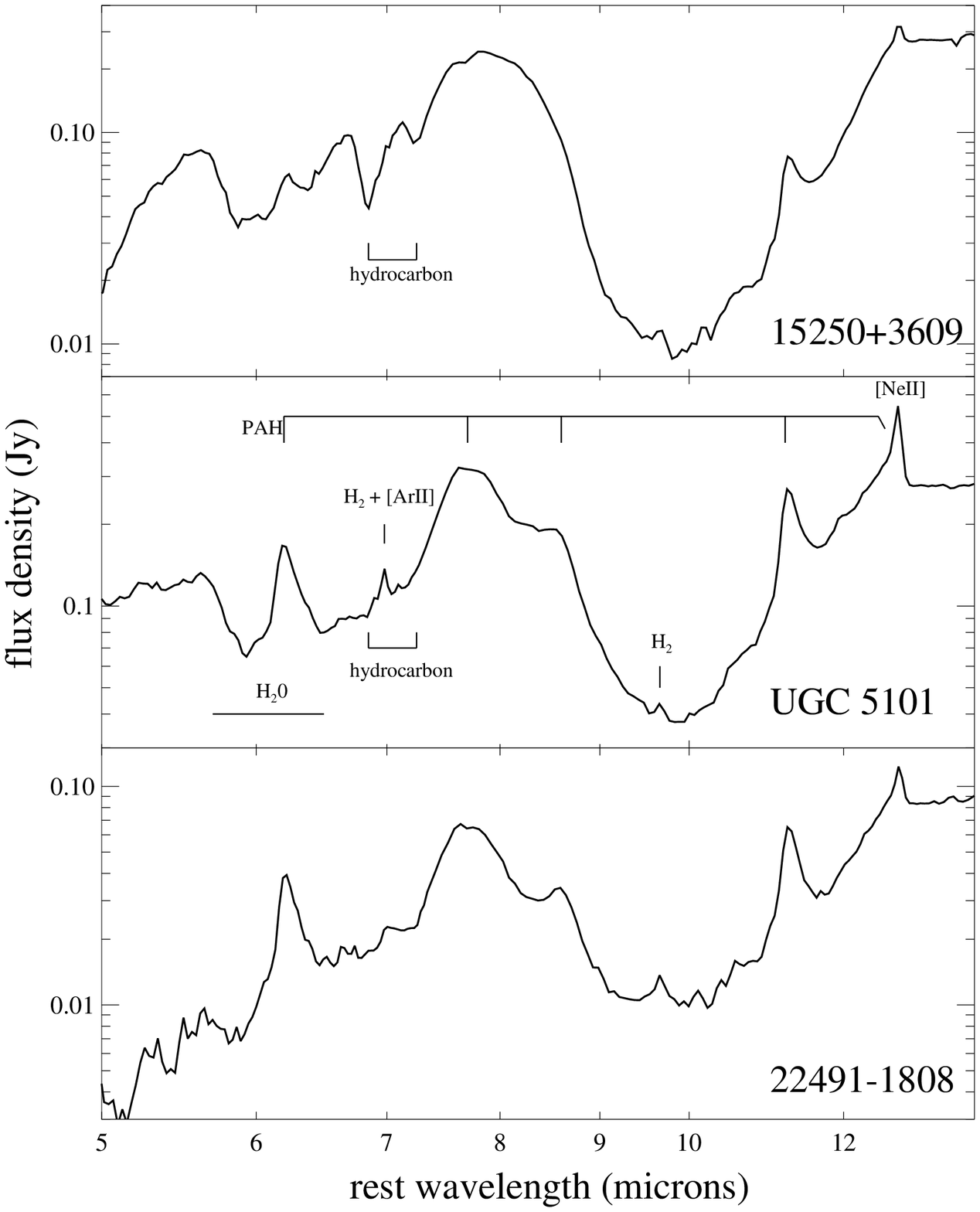}
\end{figure*}

\begin{figure*}[lowres2c]
\plotone{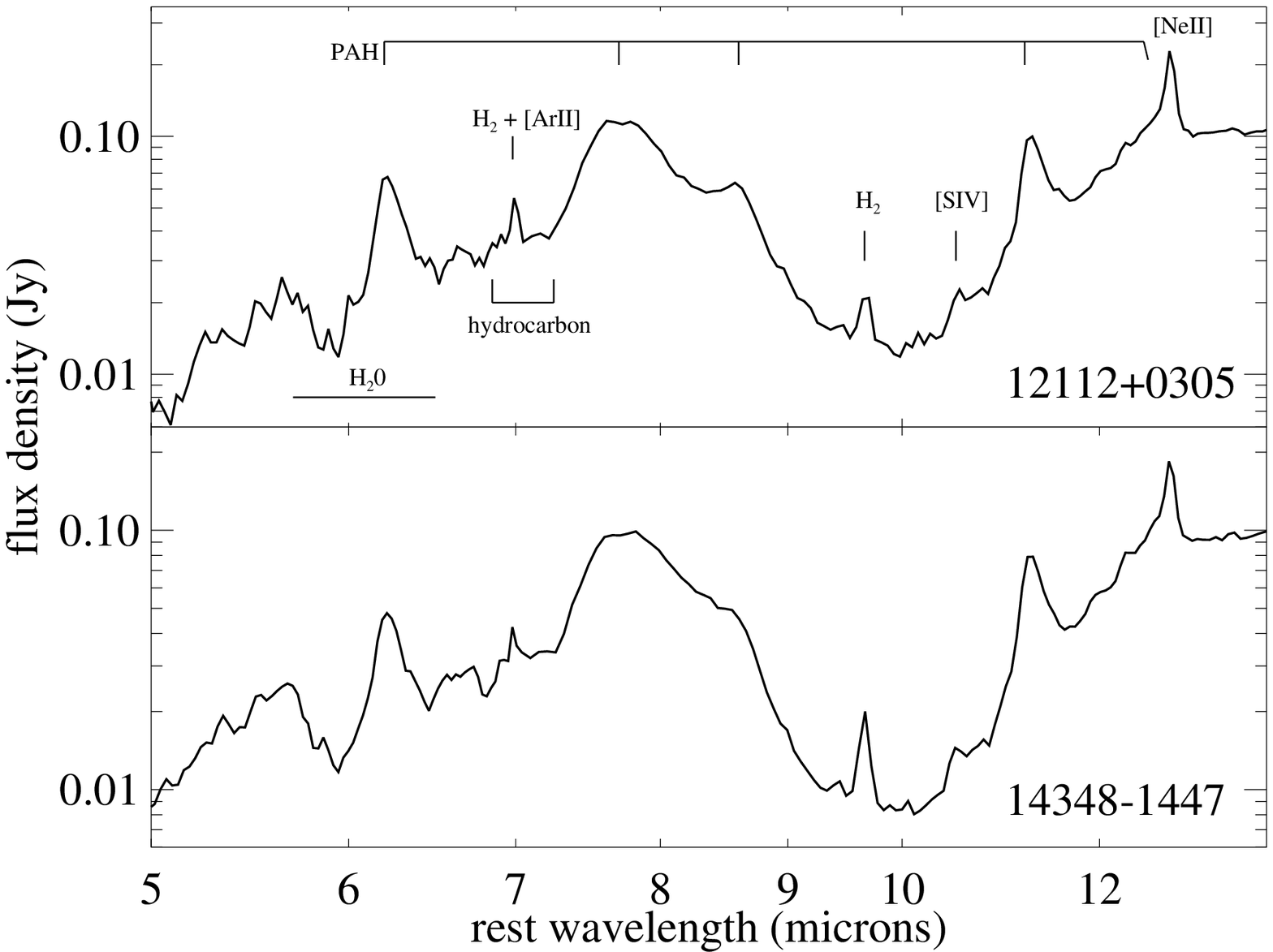}
\end{figure*}

\begin{figure*}[norm]
\plotone{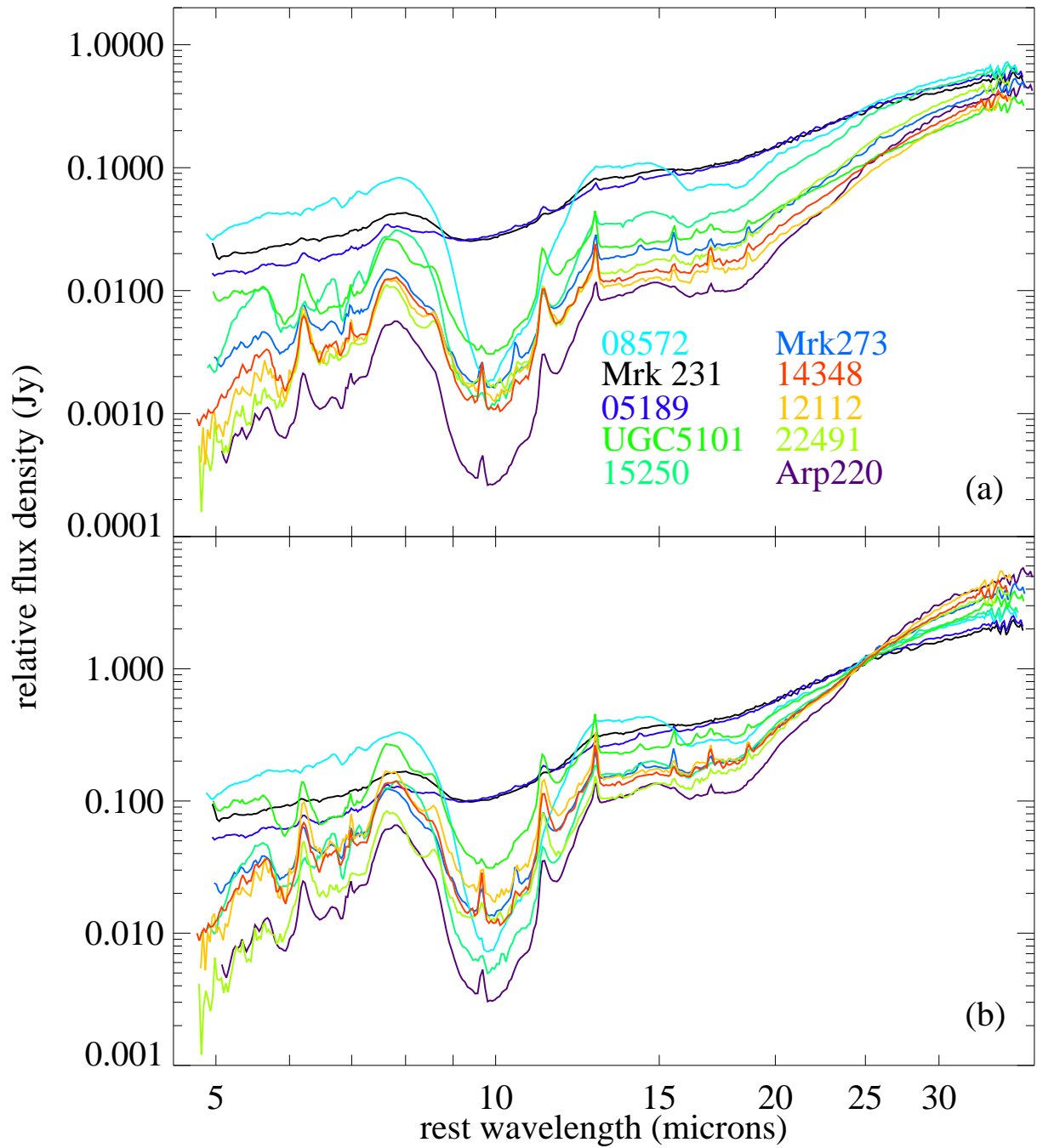}
\caption{Normalized low-resolution IRS spectra.  Spectra are normalized at
rest-frame $60\mu$m in the upper panel (a), and at rest-frame $25\mu$m in the lower
panel (b).}
\end{figure*}

\begin{figure*}[sh1]
\plotone{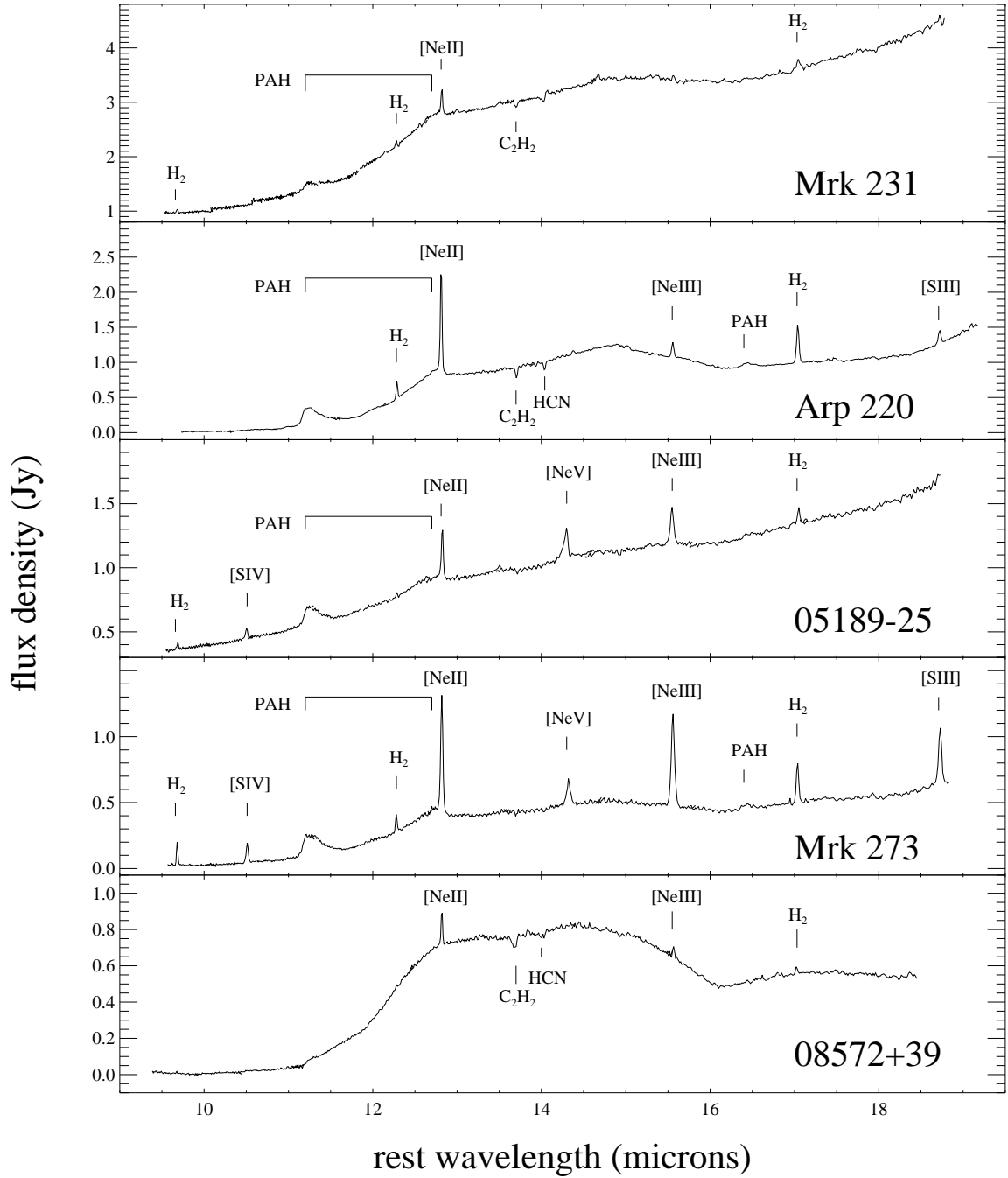}
\caption{IRS Short-High spectra of the 10 BGS ULIRGs.  Prominent
spectral features are marked.}
\end{figure*}

\begin{figure*}[sh2]
\plotone{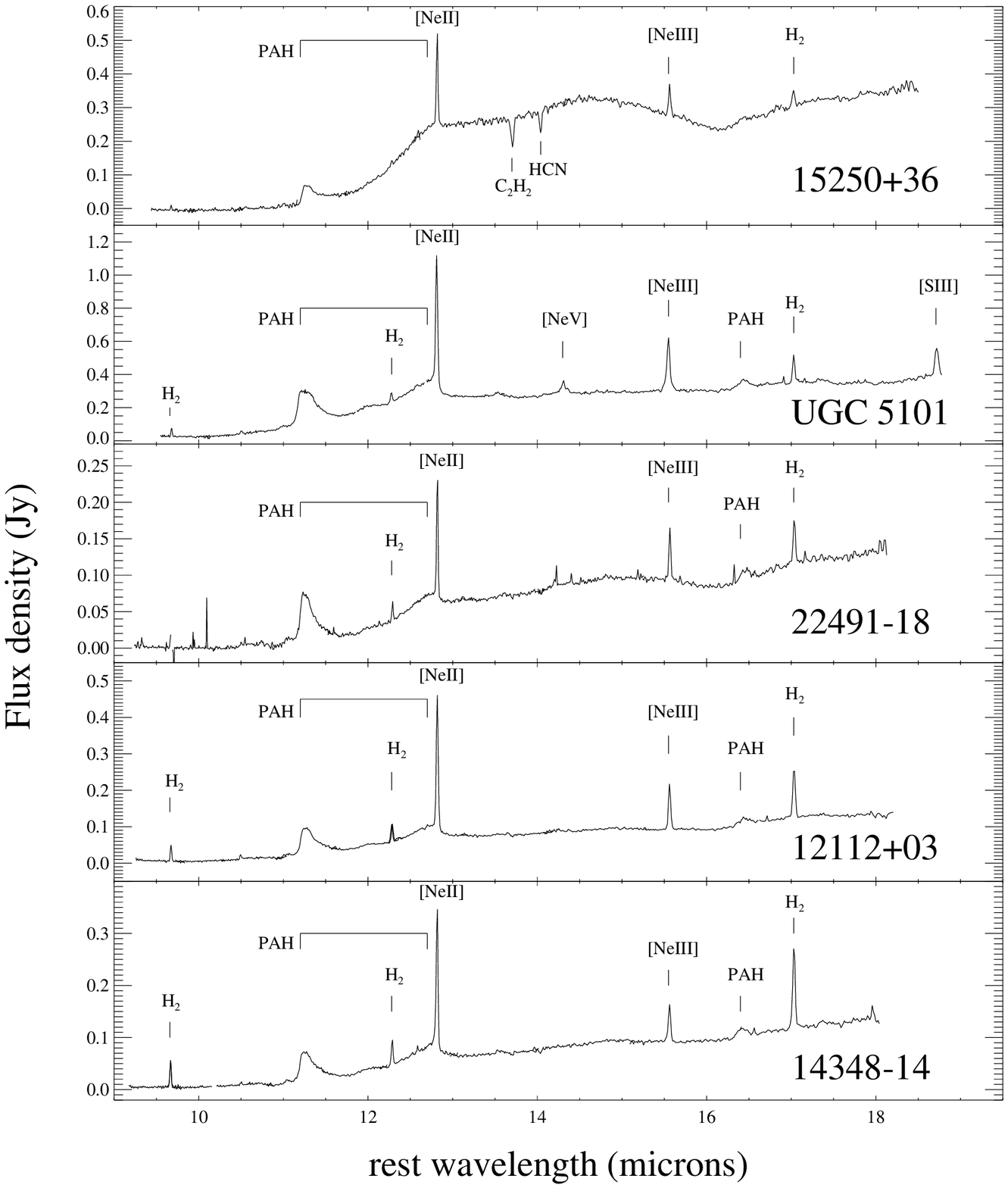}
\end{figure*}

\begin{figure*}[lh1]
\plotone{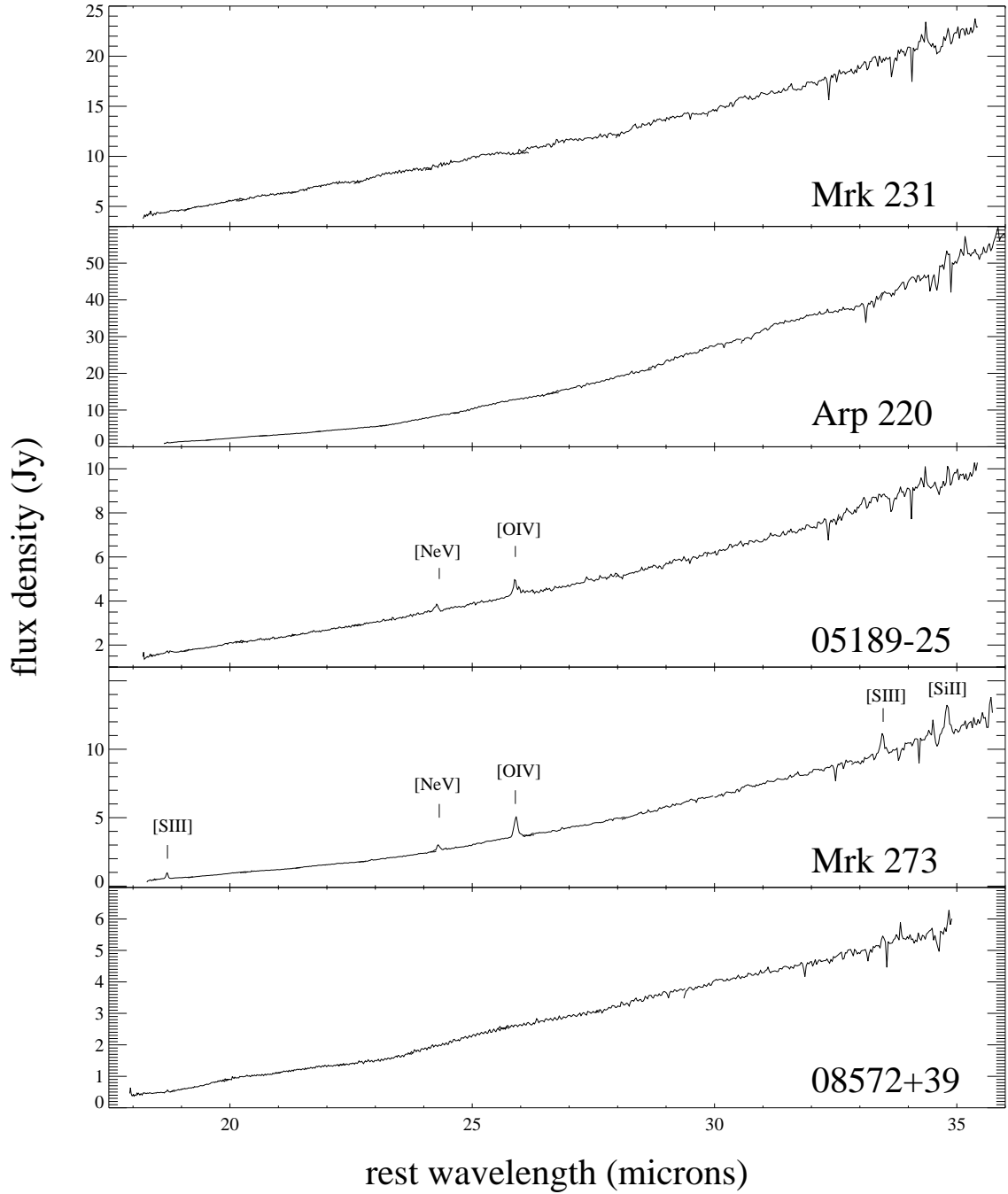}
\caption{IRS Long-High spectra of the 10 BGS ULIRGs.  Prominent
emission lines are marked.}
\end{figure*}

\begin{figure*}[lh2]
\plotone{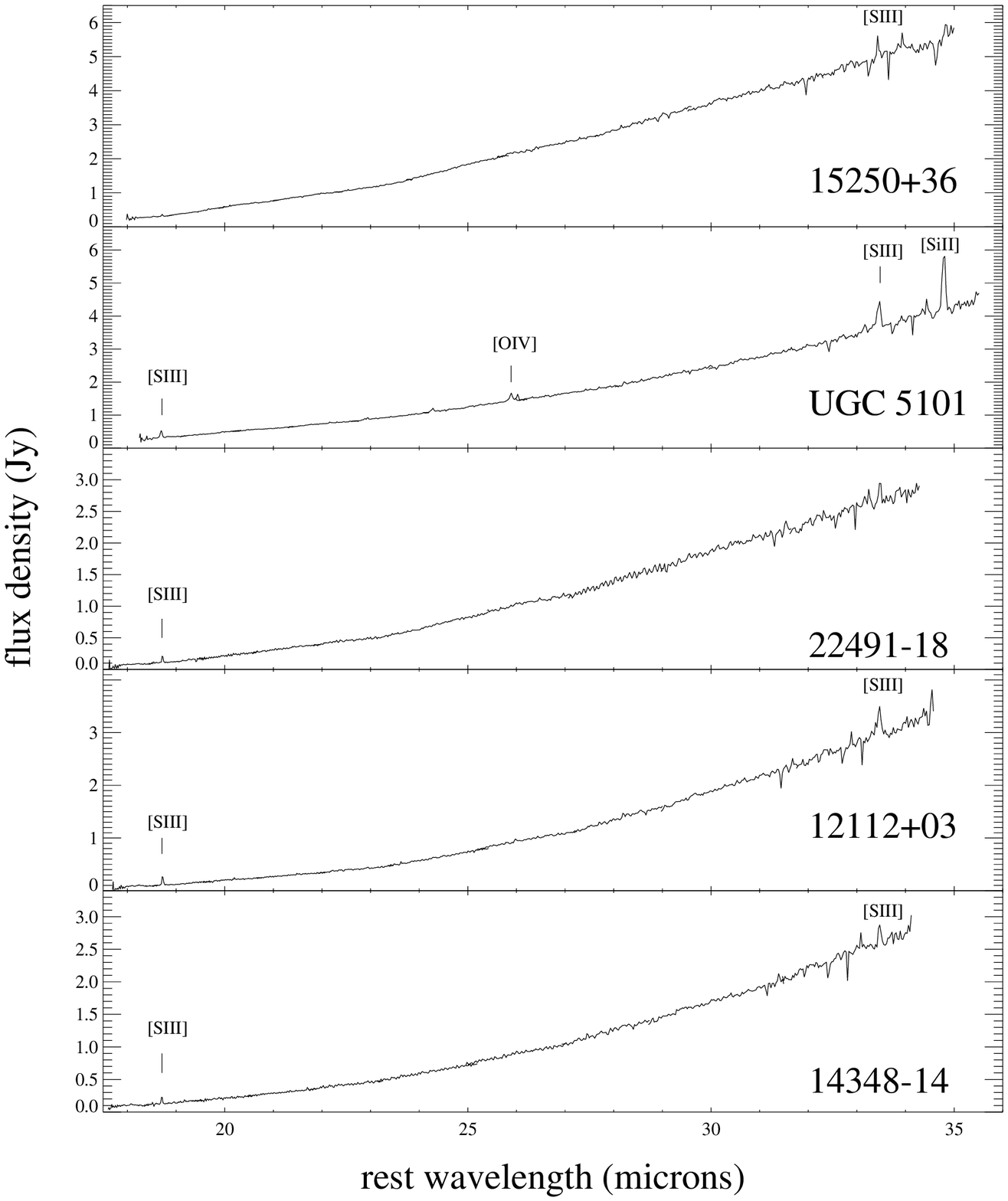}
\end{figure*}

\clearpage

\begin{figure*}[ne5ne2] 
\epsfig{file=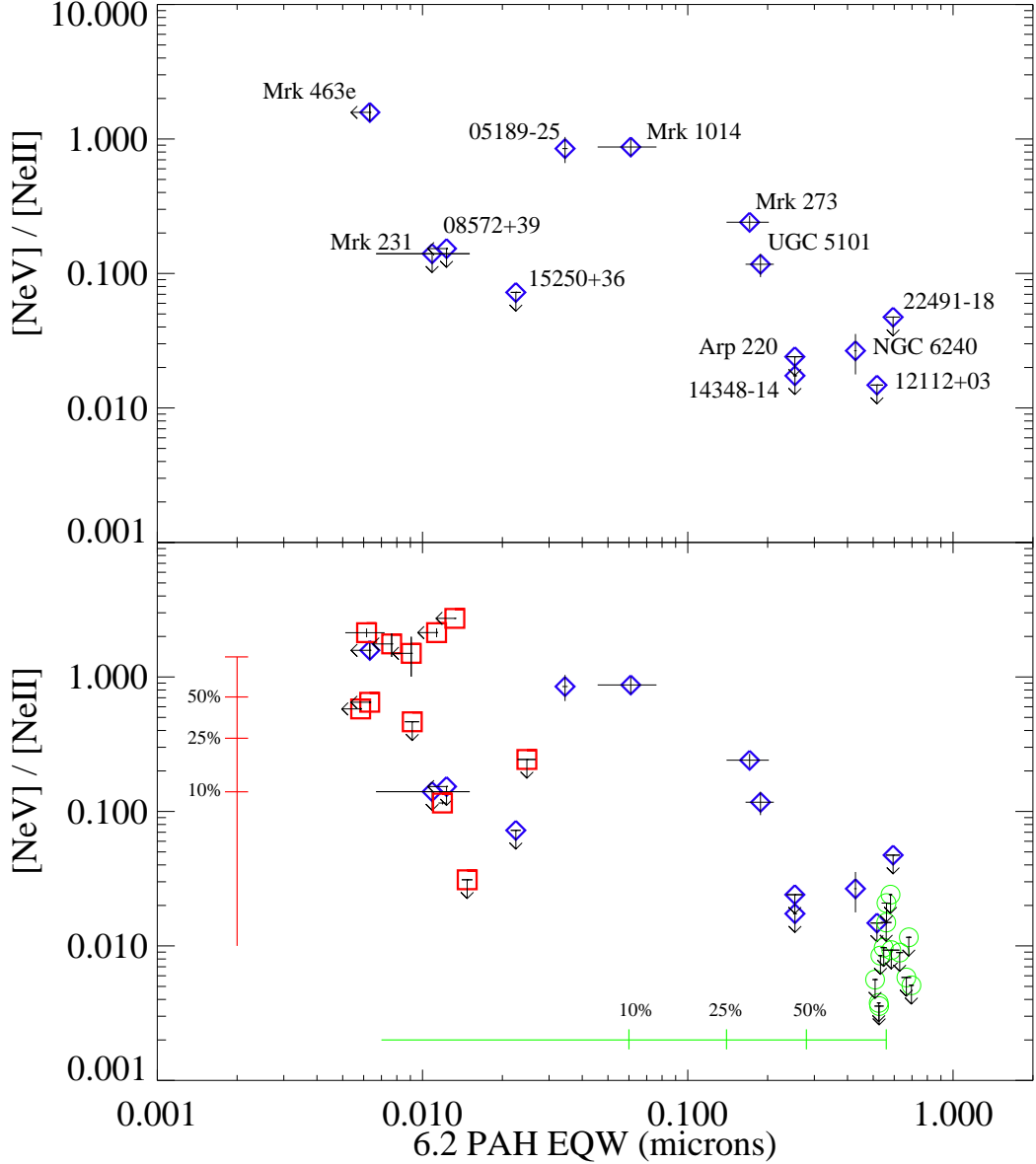,width=5.7in}
\caption{Mid-Infrared [NeV]
excitation diagram.
The [NeV] 14.3/[NeII] 12.8 vs. 6.2 PAH EQW (Fig. 5a)
for the 10 BGS ULIRGs (blue diamonds), together with a sample of AGN
(red squares - from Weedman et al. 2005) and starburst galaxies
(green circles - from Devost et al. 2006, and Brandl et al. 2006).
Also plotted are
the ULIRGs Mrk 1014 and Mrk 463e (from Armus et al. 2004) and NGC 6240
(from Armus et al. 2005).  The upper panel shows only the ULIRGs (labeled)
while the lower panel shows the ULIRGs together with the AGN and
Starburst galaxies.
In all cases, one-sigma error bars and upper limits are indicated.  The vertical
(red) and horizontal (green) lines indicate the fractional AGN and starburst 
contributions to the [NeV]/[NeII] and $6.2\mu$m PAH EQW, respectively, assuming 
a simple linear mixing model.  In each case, 
the 50\%, 25\% and 10\% levels are marked.  The 100\% level is set by the average 
detected values for the [NeV]/[NeII] and $6.2\mu$m EQW among the AGN and starbursts, 
respectively (see text for details).  The comparison AGN are I Zw 1, NGC 1275, Mrk 3, 
PG0804+761, PG1119+120, NGC 4151, PG1211+143, 3C 273, Cen A, Mrk 279, PG1351+640, Mrk 841
and PG2130+099.  The comparison starburst galaxies are NGC 660, NGC 1222, IC342, NGC 1614, 
NGC 2146, NGC 3256, NGC 3310, NGC 4088, NGC 4385, NGC 4676, NGC 4818, NGC 7252 and NGC 7714.}
\end{figure*}

\begin{figure*}[o4ne2]
\epsfig{file=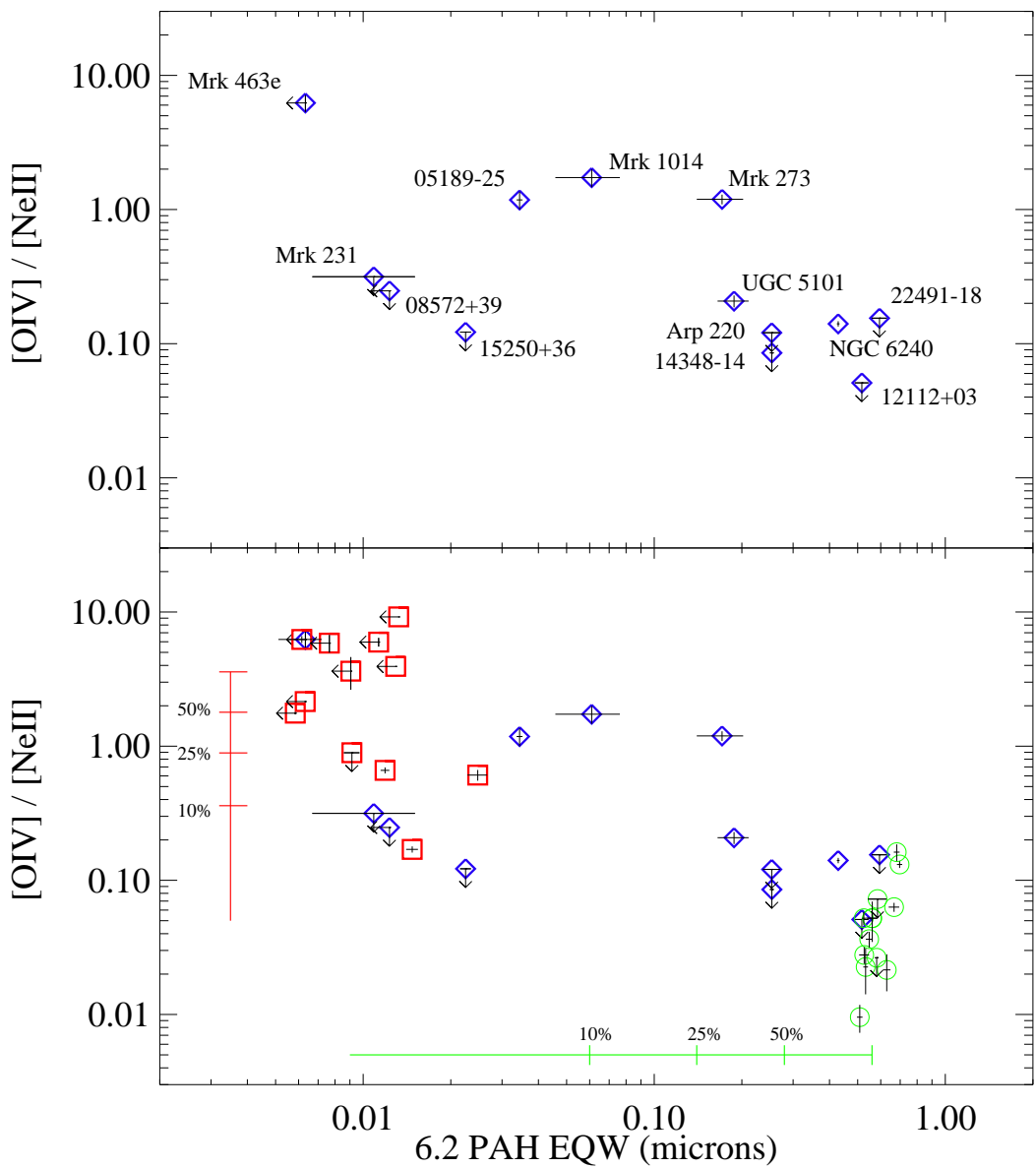,width=5.7in}
\caption{Mid-Infrared [OIV] excitation diagram.  The abcissa is the 
same as in previous figure, but the ordinate is now the [OIV] 25.9/[NeII] 12.8 line
flux ratio.  All points are defined as in the previous figure.  Fractional AGN and 
starburst contributions are indicated as in the previous figure.}
\end{figure*}

\begin{figure*}[ne5toFIR] 
\epsfig{file=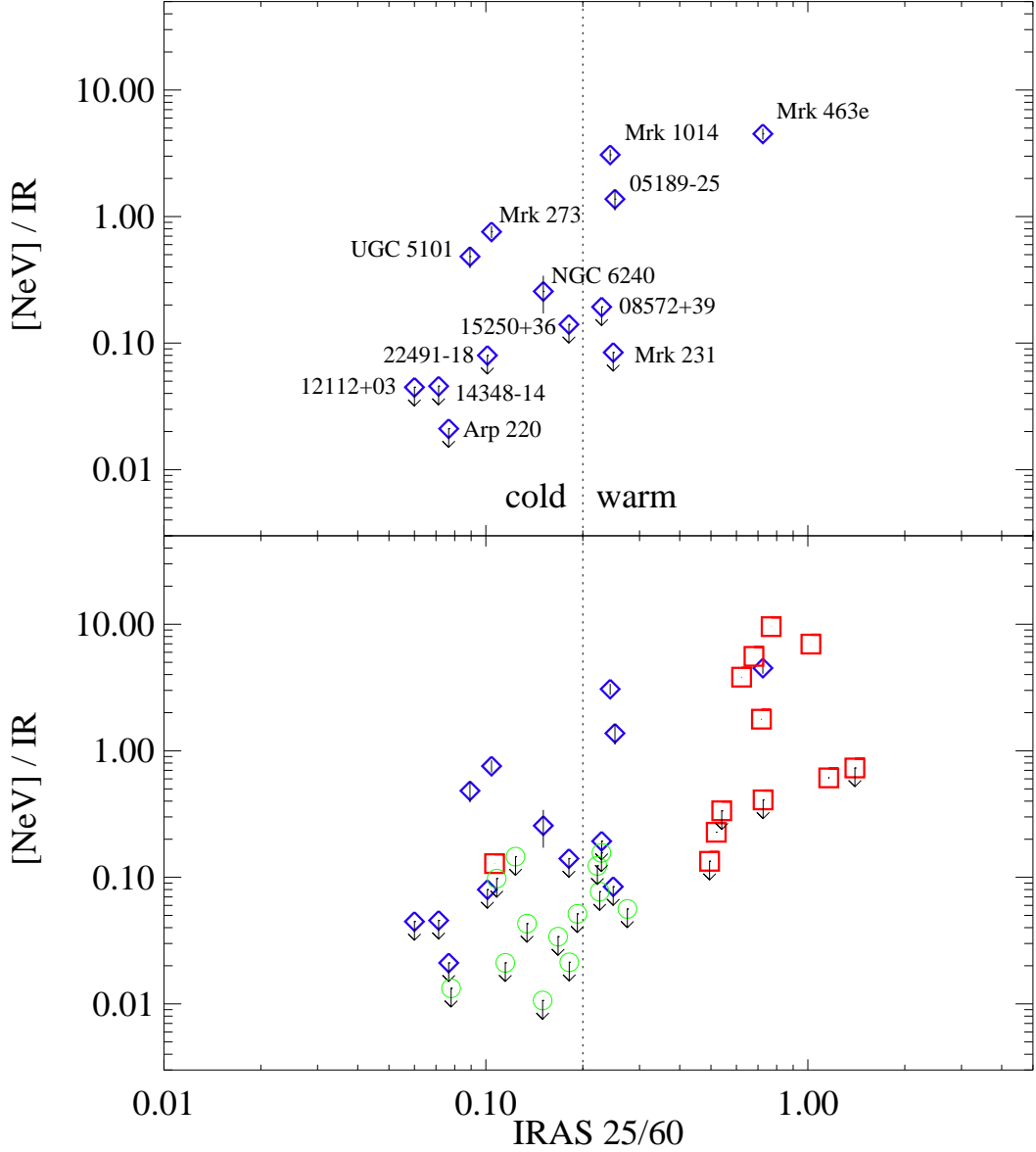,width=5.7in}
\caption{Observed
[NeV] $14.3\mu$m line flux, scaled to total infrared ($8-1000\mu$m) flux vs. the IRAS 25 to 60
micron flux density ratio for the same sources as in Fig.5.  The vertical 
axis is in units of $10^{-4}$.  The upper panel
shows the ULIRGs alone, while the lower panel shows the ULIRGs
together with the AGN and Starburst galaxies.  The dotted vertical
line indicates the dividing line between ``cold" (25/60 $< 0.2$) and ``warm" (25/60 $> 0.2$) 
IRAS sources.} 
\end{figure*}

\begin{figure*}[laurent_mix]
\epsfig{file=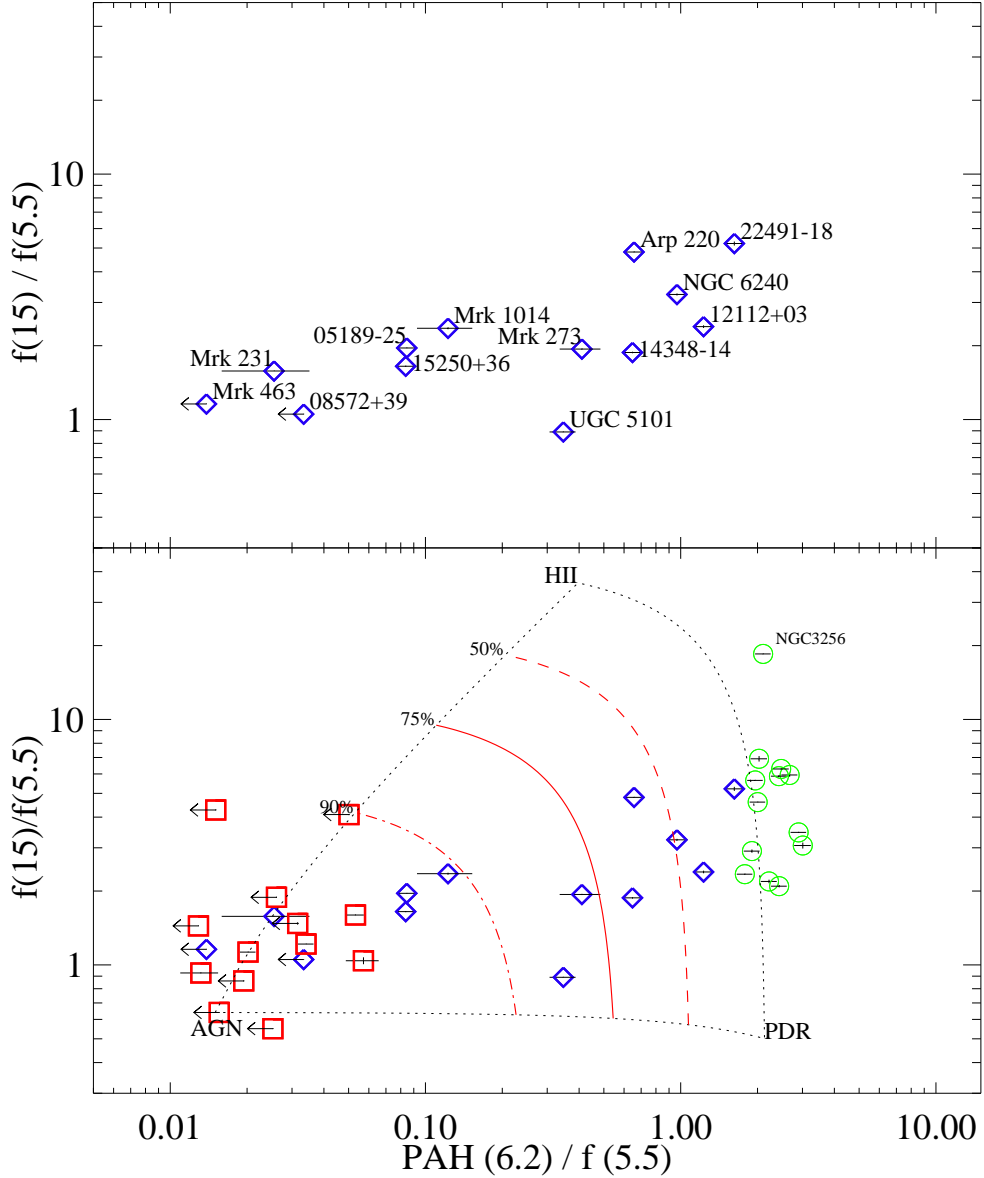,width=5.7in}
\caption{Mid-Infrared diagnostic diagram (Laurent et al. 2000) comparing the integrated continuum 
flux from $14-16\mu$m
(f15), the integrated continuum flux from $5.3-5.8\mu$m (f5.5), and the $6.2\mu$m
PAH flux.  All points are defined as in Fig.5.  The upper panel shows the 
ULIRGs (labelled), while the lower
panel shows the ULIRGs together with the AGN and
Starburst galaxies. The three vertices, labelled as AGN, HII and PDR, represent
the positions of 3C273 (Weedman et al. 2005), M17, and NGC 7023 (Peeters et al. 2003). 
In all cases, one-sigma error bars and upper limits are indicated.
The red lines indicate a 90\% (dot-dashed), 75\% (solid), and 50\% (dot-dashed) 
fractional AGN contribution, where 100\% AGN is defined by the 
position of 3C273 (see text for details).}
\end{figure*}

\begin{figure*}[14348_sedfits]
\plotone{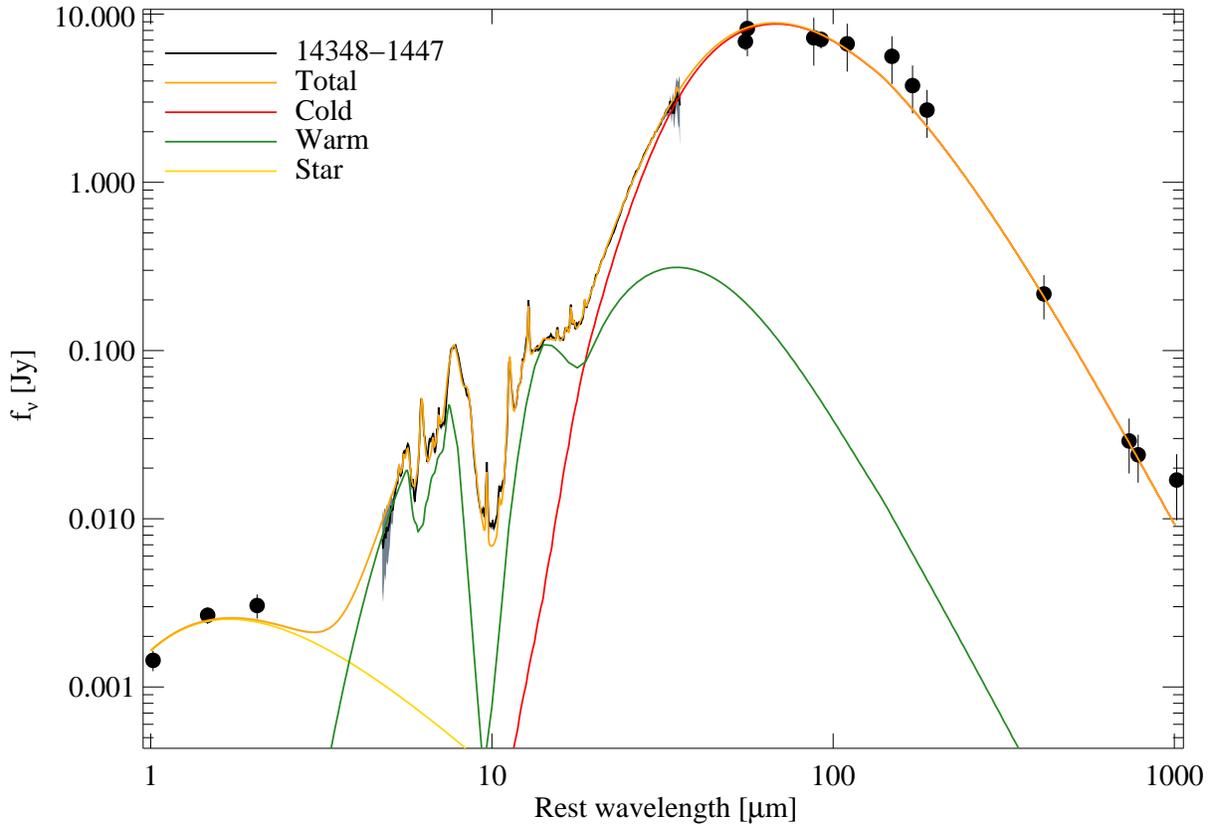}
\caption{SED fits of the BGS ULIRGs.  While all sources are fit with dust, 
PAH, and Gaussian components, only the stellar and 
continuum dust components ($2-3$ depending upon the source) are shown here, 
color-coded for clarity.  In all cases, 
the light gray
envelope is $1\sigma$ uncertainty of the total fit.}
\end{figure*}

\begin{figure*}[sedfits]
\plotone{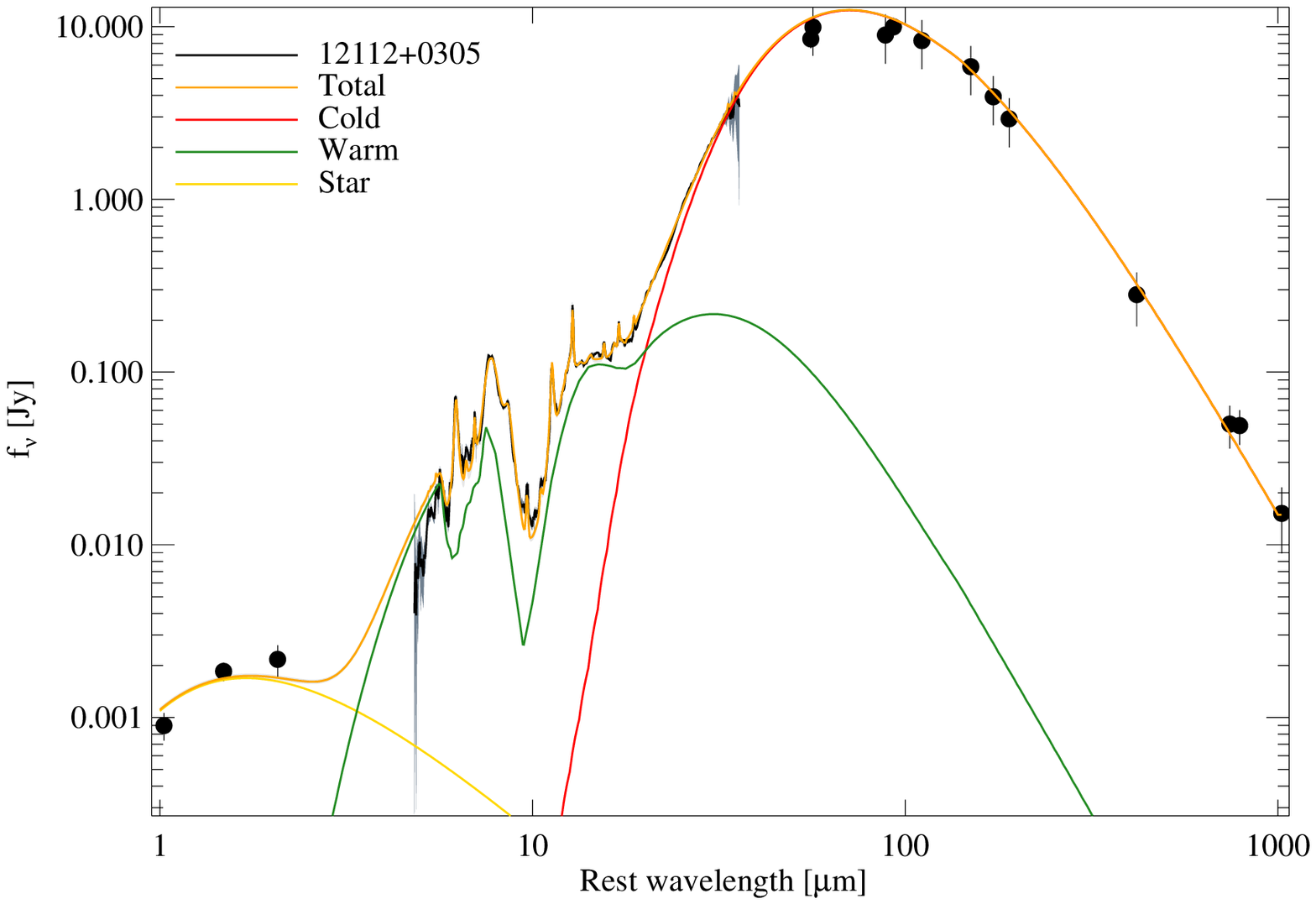}
\plotone{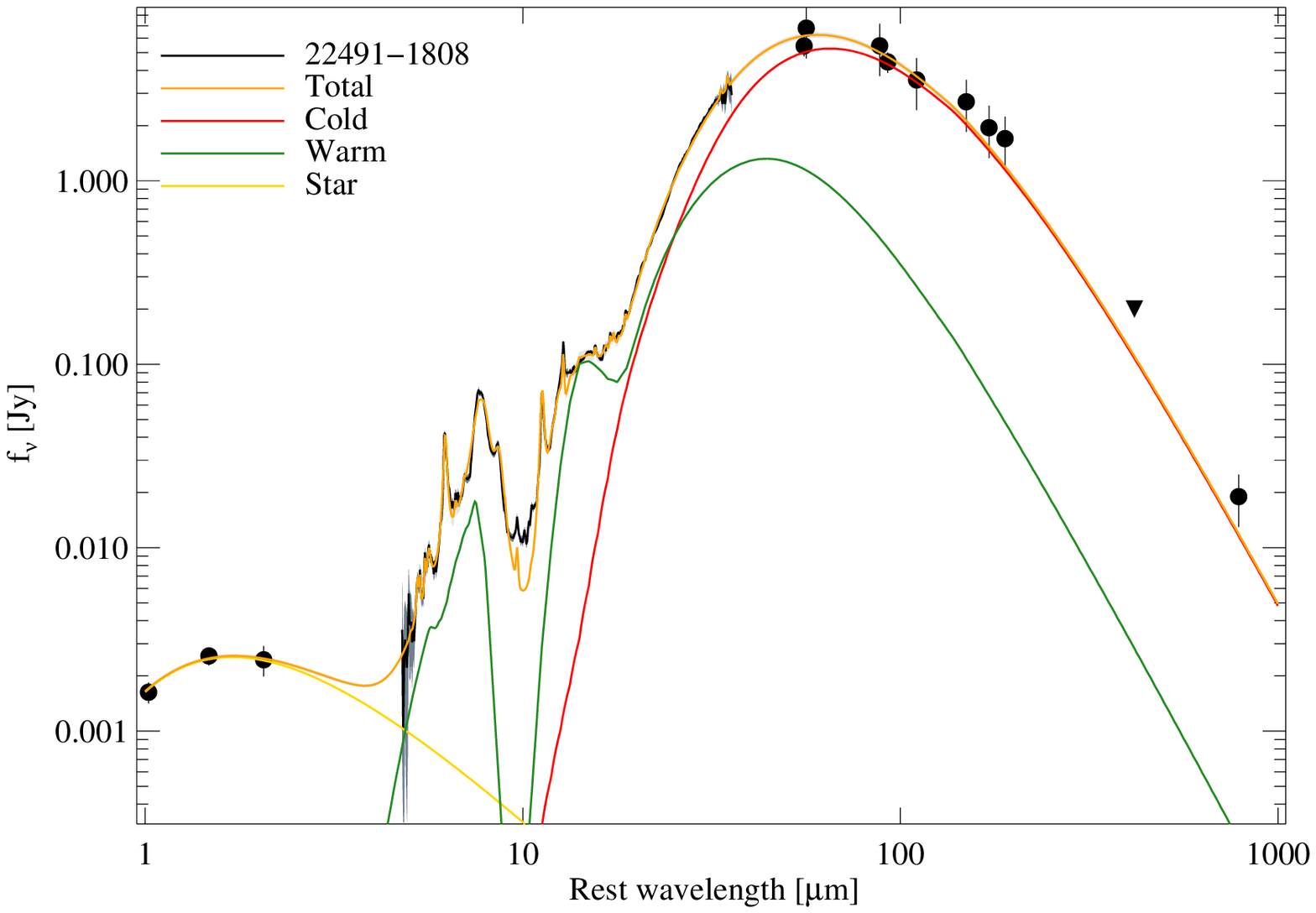}
\end{figure*}

\begin{figure*}[sedfits]
\plotone{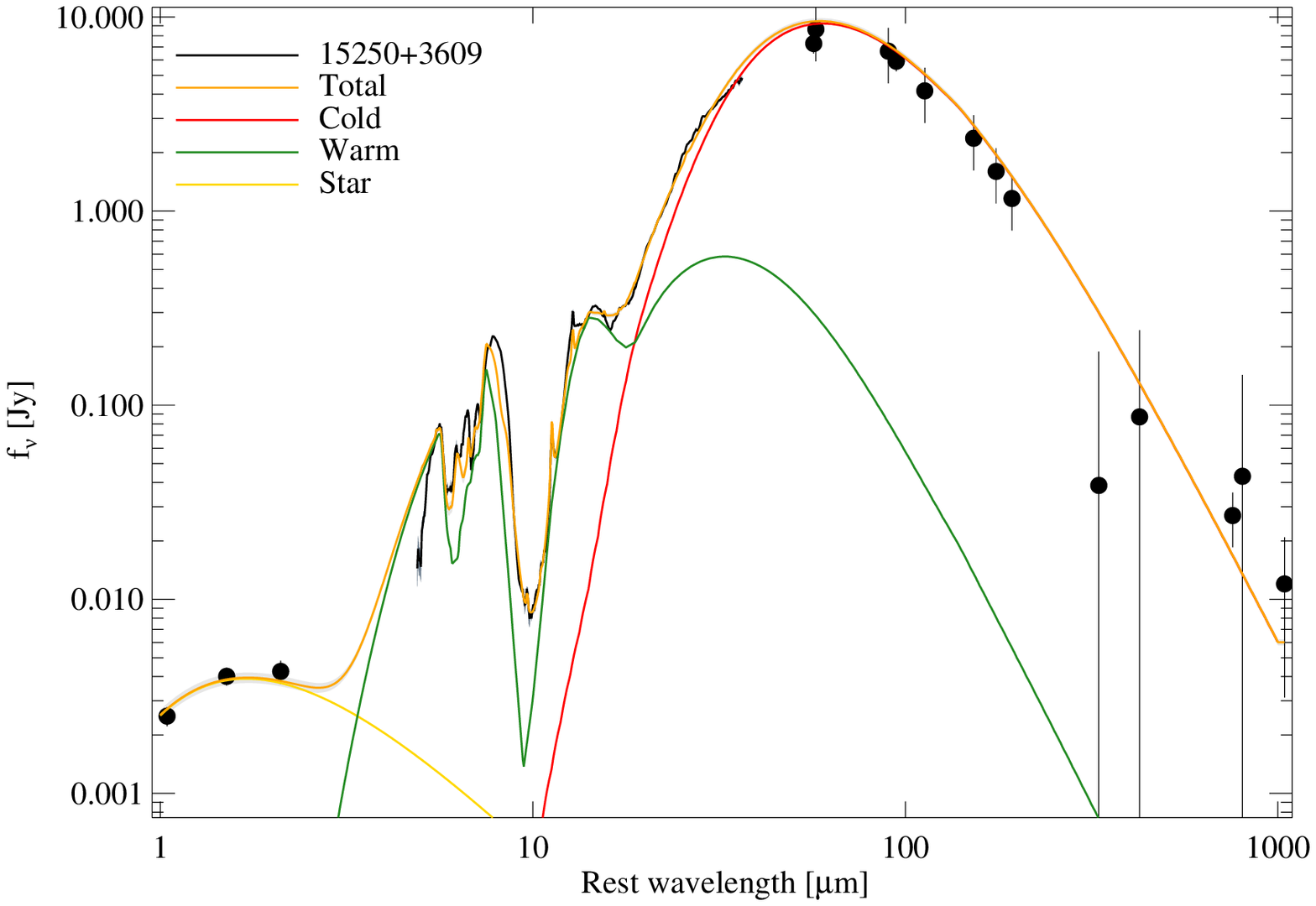}
\plotone{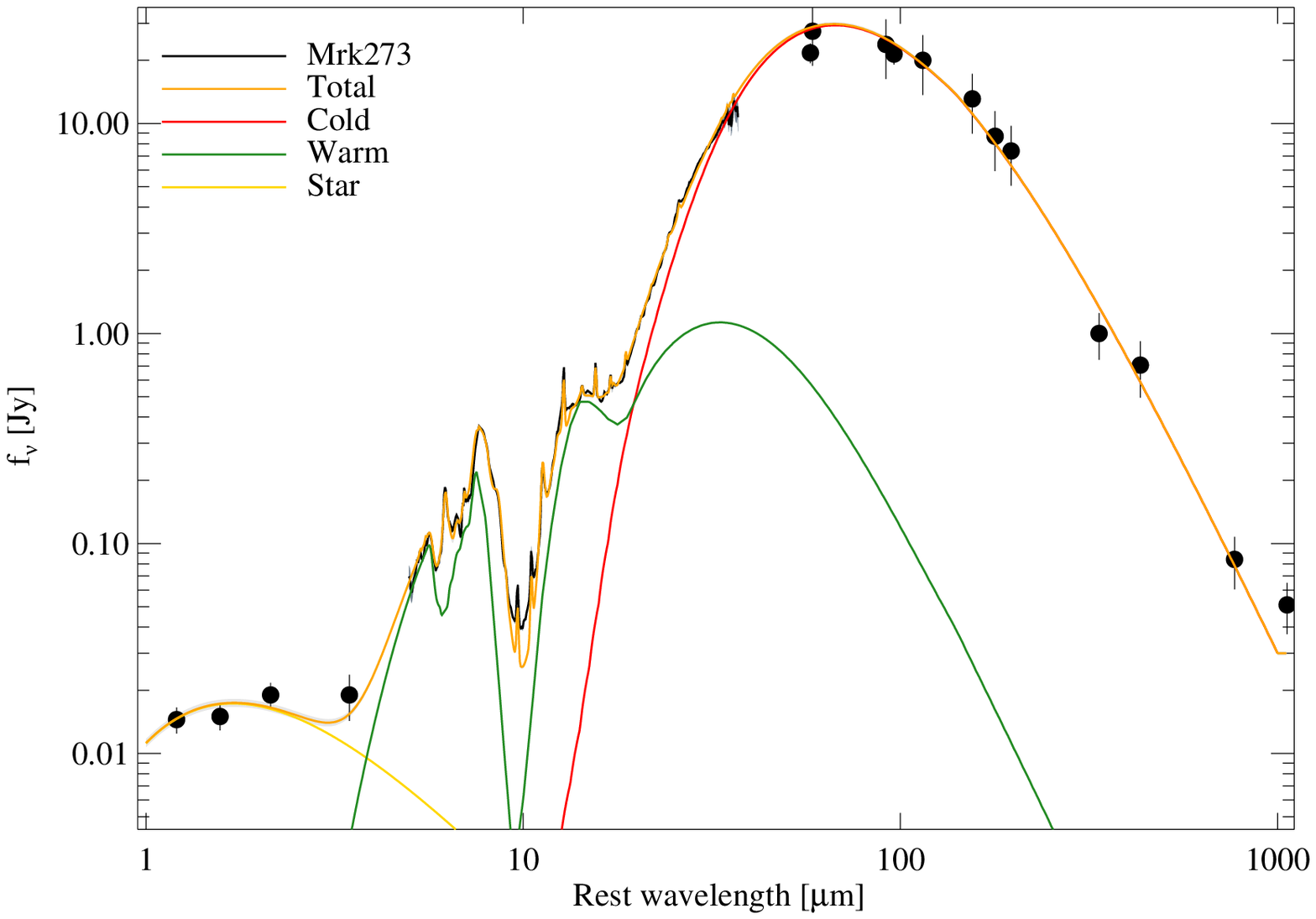}
\end{figure*}

\begin{figure*}[sedfits]
\plotone{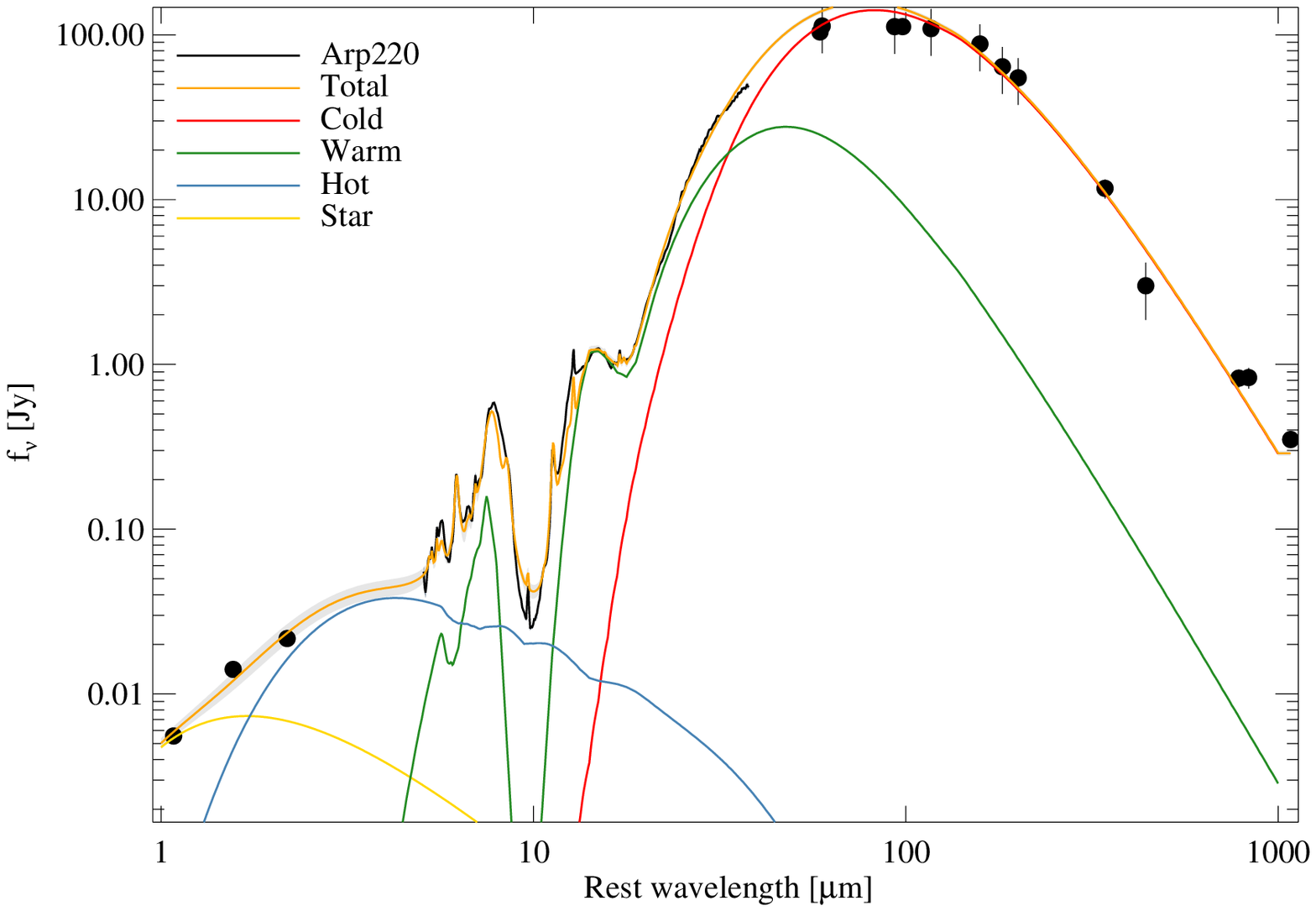}
\plotone{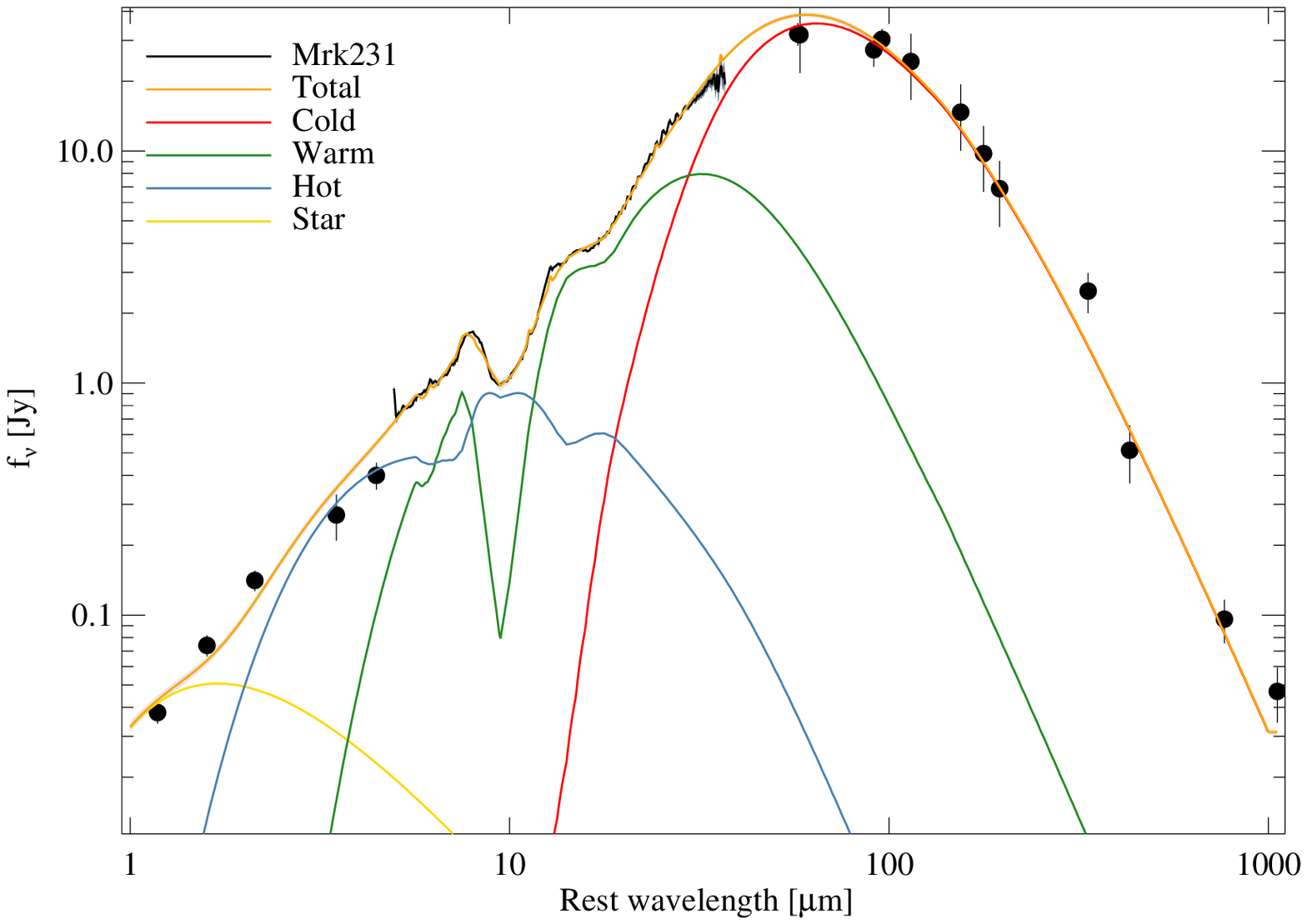}
\end{figure*}

\begin{figure*}[sedfits]
\plotone{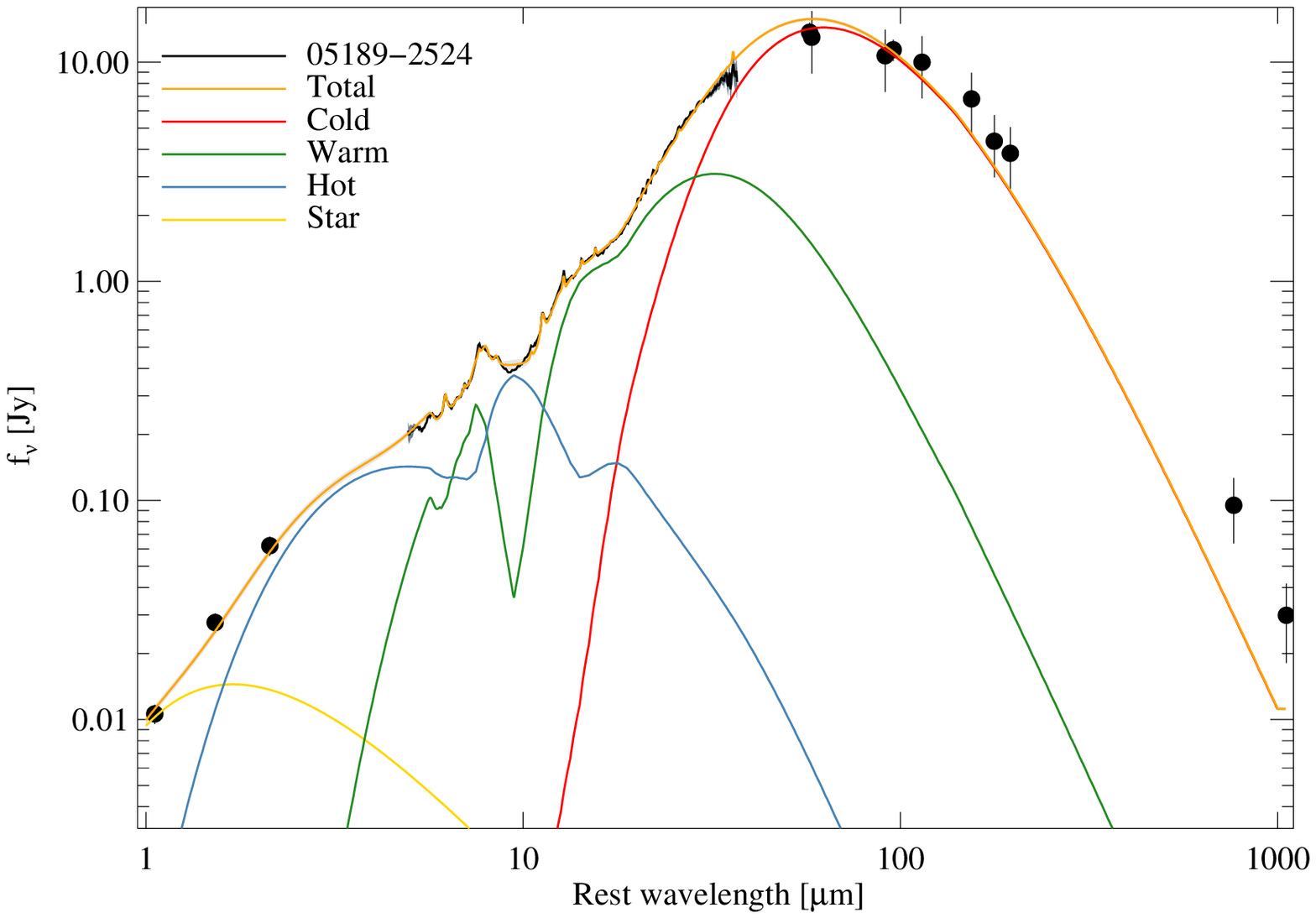}
\plotone{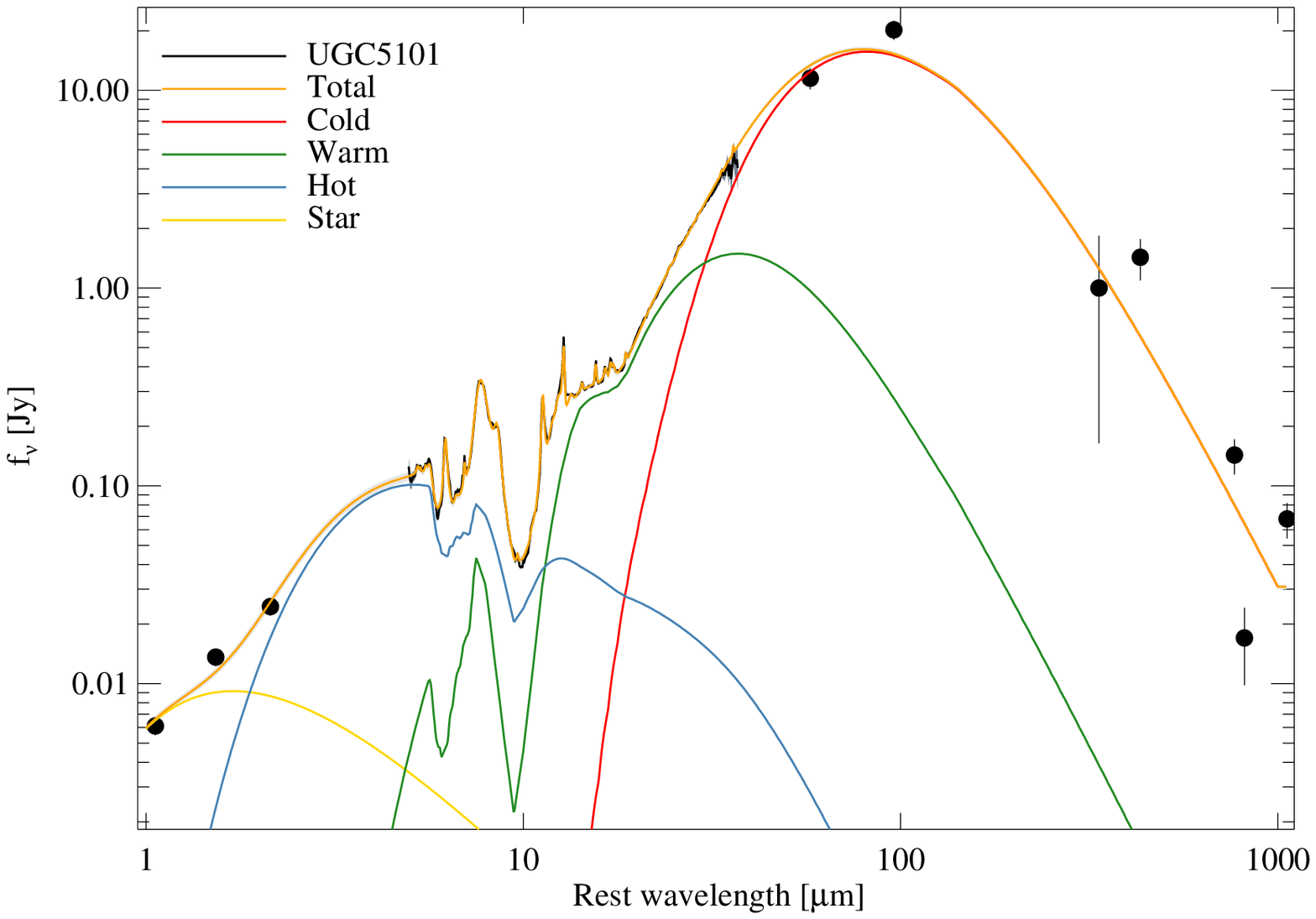}
\end{figure*}

\begin{figure*}[sedfits]
\plotone{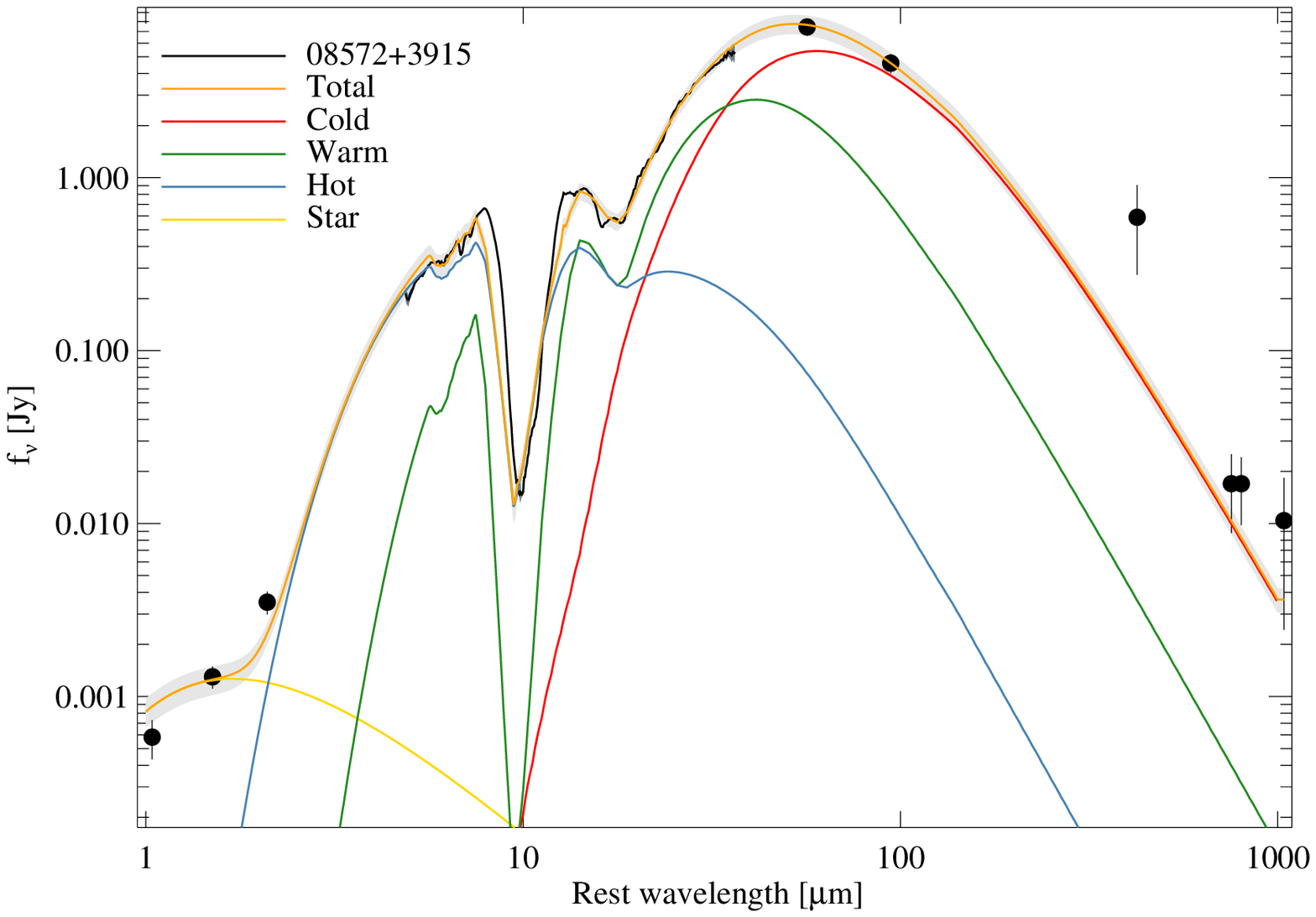}
\end{figure*}

\begin{figure*}[8_24]
\epsfig{file=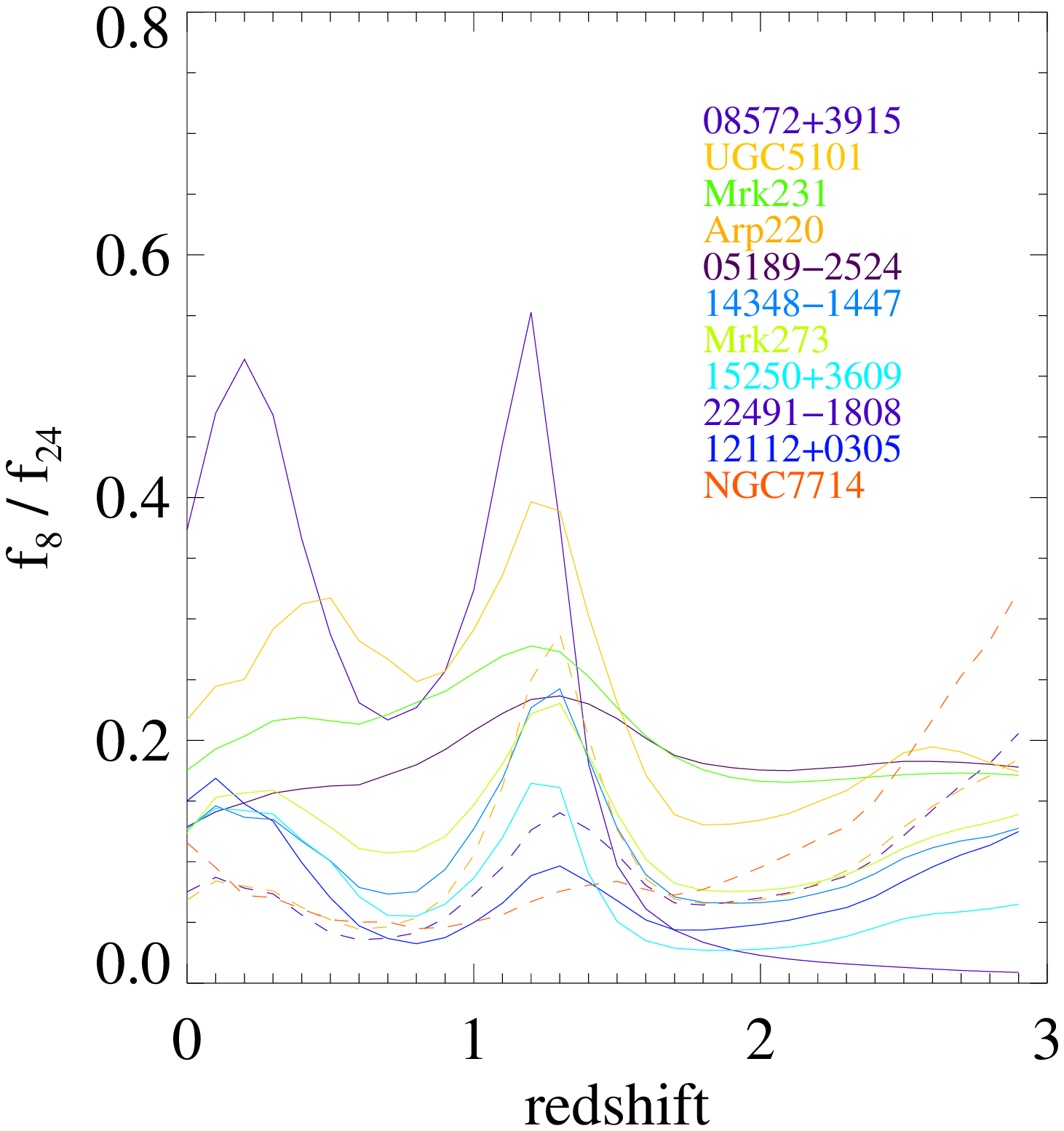,width=5.7in}
\caption{Mid-Infrared IRAC 8-micron to MIPS $24\mu$m filter flux density ratio
as a function of redshift for the BGS ULIRGs,
and the nearby starburst galaxy, NGC
7714 (Brandl et al. 2004).  The galaxies are listed in the legend from top to bottom
in (decreasing) order of their 8/24 ratio at $z=1.3$, with IRAS 08572+3915
having the largest ratio and NGC 7714 having the smallest ratio.  Arp 220,
IRAS 22491-1808, and NGC 7714 are drawn with dashed lines for clarity.}
\end{figure*}

\begin{figure*}[16_24]
\epsfig{file=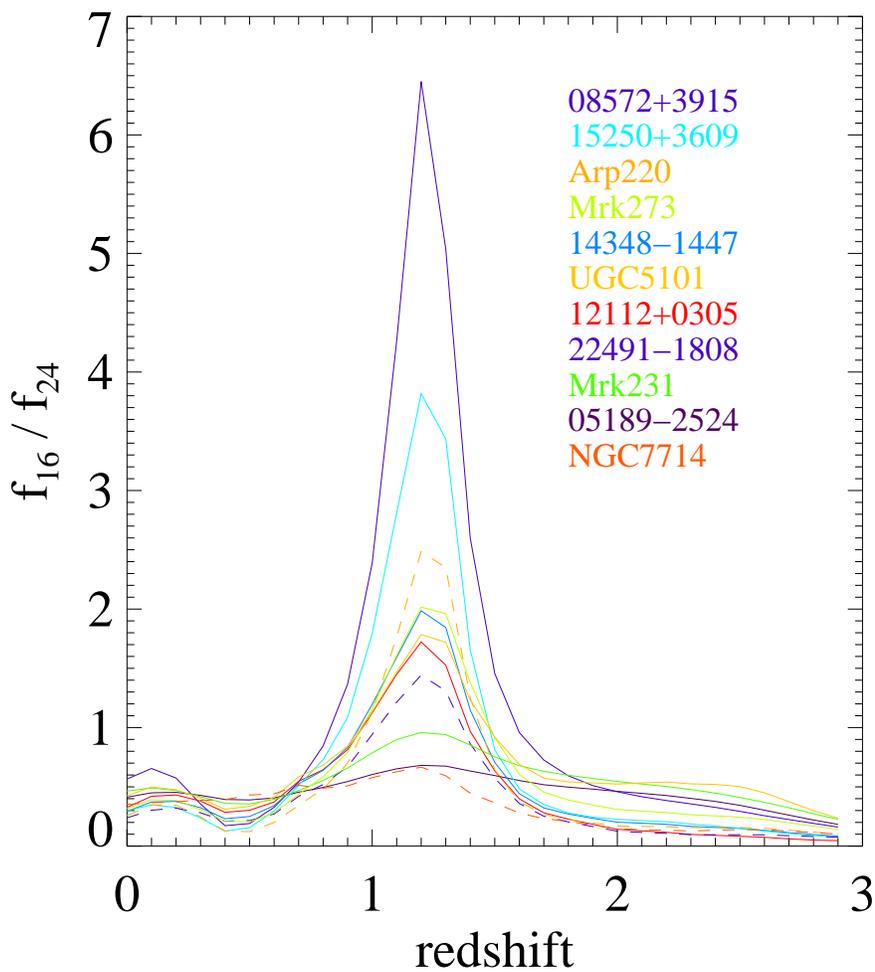,width=5.7in}
\caption{Mid-Infrared IRS Blue Peakup to MIPS $24\mu$m filter flux density ratio
as a function of redshift for the BGS ULIRGs,
and the nearby starburst galaxy, NGC
7714 (Brandl et al. 2004).  The galaxies are listed in the legend from top to bottom
in (decreasing) order of their 16/24 ratio at $z=1.3$, with IRAS 08572+3915
having the largest ratio and NGC 7714 having the smallest ratio.  Arp 220,
IRAS 22491-1808, and NGC 7714 are drawn with dashed lines for clarity.}
\end{figure*}

\clearpage
\references

\reference{} Alonso-Herrero, A., Quillen, A.C., Simpson, C., Efstathiou, A., \& Ward, M.J. 2001, AJ, 121, 1369.

\reference{} Armus, L., Heckman, T.M, \& Miley, G.K. 1987, AJ, 94, 831.

\reference{} Armus, L., Heckman, T.M., \& Miley, G.K. 1989, 347, 727.

\reference{} Armus, L. et al. 2004, ApJ Suppl., 154, 178.

\reference{} Armus, L., et al. 2006, ApJ, 640, 204.

\reference{} Benford, D.J. 1999, Ph.D. Thesis, California Institute of Technology.

\reference{} Bernard-Salas, J., Pottasch, S.R., Beintema, D.A., \& Wesselius, P.R. 2001, Astron. \& Astrophys., 367, 949.

\reference{} Boonman, A.M.S., et al. 2003, Astron. \& Astrophys., 406, 937.

\reference{} Brandl, B.R., et al. 2004, ApJ Suppl., 154, 188.

\reference{} Brandl, B.R., et al. 2006, ApJ, in press.

\reference{} Charmandaris, V., et al. 2002, Astron. \& Astrophys., 391, 429.

\reference{} De Grijp, M.H.K., Miley, G.K., Lub, J., De Jong, T. 1985, Nature, 314, 240.

\reference{} Devost, D., et al. 2006, in preparation.

\reference{} Dudley, C.C., \& Wynn-Williams, C.G. 1997, ApJ, 488, 720.

\reference{} Genzel, R., Lutz, D., Sturm, E., Egami, E., Kunze, D., 
et al. 1998, ApJ, 498, 589.


\reference{} Higdon, S.J.U., et al. 2004, PASP, 116, 975.

\reference{} Higdon, S.J.U., et al. 2006, ApJ, in press.

\reference{} Houck, J.R., et al. 2004, ApJ Suppl., 154, 18.

\reference{} Houck, J.R. et al. 2005, ApJ, 622, 105L.


\reference{} Imanishi, M., Dudley, C.C., \& Maloney, P.R. 2001, ApJ, 558, L93.

\reference{} Imanishi, M., Dudley, C.C., \& Maloney, P.R. 2006, ApJ, 637, 114.

\reference{} Imanishi, M., Terashima, Y., Anabuki, N., \& Nakagawa, T. 2003, ApJ, 596, L167.

\reference{} IRAS Catalogs and Atlases: Point Source Catalog. 1985 (Washington, DC; US Government Printing Office).

\reference{} Kasliwal, M.M., et al. 2005, ApJ, 634, 1L.

\reference{} Kim, D.C., \& Sanders, D.B. 1998, ApJS, 119, 41.

\reference{} Klaas, U., et al. 2001, Astron. \& Astrophys., 379, 823.

\reference{} Lahuis, F., \& van Dishoeck, E.F. 2000, Astron. \& Astrophys., 355, 699.

\reference{} Laurent, O., et al. 2000, A \& A, 359, 887.


\reference{} Lutz, D., Kunze, D., Spoon, H.W.W., \& Thornley, M.D. 1998, 
A \& A, 333, L75.

\reference{} Lutz, D., Veilleux, S., \& Genzel, R. 1999, ApJ, 517, L13.

\reference{} Lutz, D., et al. 2003, Astron. \& Astrophys., 409, 867.

\reference{} Maiolino, R., Krabbe, A., Thatte, N., \& Genzel R. 1998, ApJ, 493, 650.

\reference{} Maiolino, R., et al. 2000, ApJ, 531, 219.

\reference{} MacKenty, J.W., \& Stockton, A. 1984, ApJ, 283, 64.

\reference{} Marshall, J.A., et al. 2006, ApJ, submitted.

\reference{} Mazzarella, J.M., Gaume, R.A., Soifer, B.T., Graham, J.R., Neugebauer, G.,
\& Matthews, K. 1991, AJ, 102, 1241.

\reference{} Mihos, C.J., \& Hernquist, L. 1996, ApJ, 464, 641.

\reference{} Miller, J.S. \& Goodrich, R.W. 1990, ApJ, 355, 456.

\reference{} Moshir, et al. 1990, IRAS Faint Source Catalog, V2.0.

\reference{} Moutou, C., Verstraete, L., Leger, A., Sellgren, K., \& Schmidt, W. 2000, 
A \& A, 354, L17.

\reference{} Murphy, T.W. Jr., Armus, L., Matthews, K., Soifer, B.T., Mazzarella, J.M., 
Shupe, D.L., Strauss, M.A., \& Neugebauer, G. 1996, AJ, 111, 1025.

\reference{} Peeters, E., Spoon, H.W.W., \& Tielens, A.G.G.M. 2004, ApJ, 613, 986.

\reference{} Perault, M. 1987, Ph.D. thesis, University of Paris, Paris.

\reference{} Pottasch, S.R., Beintema, D.A., Bernard-Salas, J., \& Feibelman, W.A., 2001, Astron. \& Astrophys. 380, 648.

\reference{} Prieto, M.A., Marco, O., \& Gallimore, J. 2005, astro-ph/0509181.

\reference{} Reach, W.T., Morris, P., Boulanger, F., \& Okumura, K. 2004, Icarus, in press.

\reference{} Rigopoulou, D., Kunze, D., Lutz, D., Genzel, R., \& Moorwood, 
A.F.M. 2002, A \& A, 389, 374.

\reference{} Rigopoulou, Spoon, H.W.W., Genzel, R., Lutz, D., Moorwood, 
A.F.M., \& Tran, Q.D. 1999, AJ, 118, 2625.

\reference{} Sanders, D.B., et al. 1988a ApJ, 325, 74.

\reference{} Sanders, D.B., et al. 1988b, ApJ, 328, 35L.
 
\reference{} Sanders, D.B., Soifer, B.T., Elias, J.H., Neugebauer G., \&
Matthews, K. 1988b, ApJ, 328, L35.

\reference{} Scoville, N.Z., et al. 2000, AJ, 119, 991.

\reference{} Shuder, J.M., \& Osterbrock, D.E. 1981, ApJ, 250, 55.

\reference{} Smith, J.D.T., et al. 2004, ApJ Suppl., 154, 199.

\reference{} Soifer, B.T., et al. 1987, ApJ, 320, 238.

\reference{} Soifer, B.T., et al. 2000, AJ, 119, 509.


\reference{} Spoon, H.W.W., Keane, J.V., Tielens, A.G.G.M., Lutz, D., Moorwood, A.F.M., \&
Laurent, O. 2002, A \& A, 385, 1022.

\reference{} Spoon, H.W.W., et al. 2004, ApJ Suppl., 154, 184.

\reference{} Spoon, H.W.W., et al. 2006a, 638, 759.

\reference{} Spoon, H.W.W., et al. 2006b, astro-ph/0512037.

\reference{} Stanford, S.A., Stern, D., van Breugel, W., \& De Breuck, C. 2000, ApJ Suppl., 131, 
185.

\reference{} Strauss, M.A., Huchra, J.P., Davis, M., Yahil, A., Fisher, K.B., 
\& Tonry, J. 1992, ApJS, 83, 29.

\reference{} Sturm, E., et al. 2000, A \& A, 358, 481.

\reference{} Sturm, E., et al. 2002, A \& A, 393, 821.

\reference{} Takagi, T., \& Pearson, C. 2005, MNRAS, 357, 165.

\reference{} van der Tak, F.F.S., van Dishoek, E.F., Evans, N.J., Bakker, E.J., \& Blake, G.A. 1999, ApJ, 522, 991.

\reference{} Veilleux, S., Kim, D.C., Sanders, D.B., Mazzarella, J.M. \& Soifer, B.T.
1995, ApJ Suppl., 98, 171.

\reference{} Veilleux, S., et al. 1997, ApJ, 477, 631.

\reference{} Verma, A., Lutz, D., Sturm, E., Sternberg, A., Genzel, R., \& Vacca, W. 2003, A \& 
A, 403, 829.

\reference{} Voit, G.M. 1992, ApJ, 399, 495.

\reference{} Yan, L., et al. 2005, ApJ, 628, 604.

\begin{deluxetable}{lccccccc}
  \tabletypesize{\scriptsize}
  \tablecaption{BGS ULIRGs}
  \tablewidth{0pc}
  \tablehead{
    \colhead{Object} & 
    \colhead{z}   & 
    \colhead{Distance} &
    \colhead{log(L$_{\rm IR}$)\tablenotemark{a}} &
    \colhead{SL} &
    \colhead{SH} & 
    \colhead{LL} & 
    \colhead{LH}\\
    \colhead{} & 
    \colhead{} & 
    \colhead{(Mpc)} & 
    \colhead{(L$_\odot$)} & 
    \colhead{(kpc)} & 
    \colhead{(kpc)}  & 
    \colhead{(kpc)} & 
    \colhead{(kpc)}
}
\startdata
{\bf{IRAS 12540+5708 (Mrk 231)}}         & 0.042 & 183.9 & 12.57 &3.2x6.4&4.2x10.1&9.4x18.8&9.9x19.8\\
{\bf{IRAS 15327+2340 (Arp 220)}}         & 0.018 &  77.6 & 12.16 &1.4x2.8&1.8x4.3&4x8&4.2x8.4\\
{\bf{IRAS 05189-2524}} & 0.042 & 185.7 & 12.16 &3.2x6.4&4.2x10.1&9.5x19&10x20\\
{\bf{IRAS 13428+5608 (Mrk 273)}}         & 0.038 & 164.3 & 12.15 &2.9x5.8&3.7x8.2&8.4x16.8&8.8x17.6\\
{\bf{IRAS 08572+3915}} & 0.058 & 257.6 & 12.14 &4.5x9&5.9x14.2&13.1x26.2&13.9x27.8\\
{\bf{IRAS 15250+3609}} & 0.055 & 243.9 & 12.05 &4.3x8.6&5.6x13.4&12.4x24.8&13.1x26.2\\
{\bf{IRAS 09320+6134 (UGC 5101)}}        & 0.039 & 174.3 & 12.00 &3x6&4x9.6&8.9x17.8&9.4x18.8\\
{\bf{IRAS 22491-1808}} & 0.077 & 345.9 & 12.19 &6x12&7.9x19&17.6x35.2&18.6x37.2\\
{\bf{IRAS 12112+0305}} & 0.073 & 324.2 & 12.33 &5.7x11.4&7.4x17.8&16.5x33&17.4x34.8\\
{\bf{IRAS 14348-1445}} & 0.083 & 371.7 & 12.35 &6.5x13&8.5x20.4&18.9x37.8&20x40\\

\enddata

\tablecomments{Basic properties of the BGS ULIRGs.  $^{a}$The 
$8-1000\mu$m IR luminosities are
derived from the IRAS data, using the prescription of Perault (1987).  The projected
spectral apertures (slit width by effective slit length) are given in kpc for 
all sources.  For the low-res slits, a 4-pixel effective slit length is given, corresponding
to the blue end of the extraction aperture which expands with wavelength - see text
for details.}

\end{deluxetable}

\begin{deluxetable}{llccccccc}
\tabletypesize{\scriptsize}
\tablecaption{IRS Observation Log\label{tbl-1}}
\tablewidth{0pc}
\tablehead{
\colhead{Object} &
\colhead{date} &
\colhead{peakup} &
\colhead{SL1} &
\colhead{SL2} &
\colhead{LL1} &
\colhead{LL2} &
\colhead{SH} &
\colhead{LH}
}
                                                                                                              
\startdata
{\bf{Mrk 231}}&14 April 2004&BPU-2MASS&2x14&2x14&5x6&5x6&6x30&4x60\\
{\bf{Arp 220}}&29 February 2004&BPU-2MASS&3x14&3x14&5x6&5x6&6x30&4x60\\
{\bf{IRAS 05189-2524}}&22 March 2004&BPU-2MASS&3x14&3x14&2x14&2x14&6x30&4x60\\
{\bf{Mrk 273}}&14 April 2004&BPU-2MASS&2x14&2x14&2x14&2x14&6x30&4x60\\
{\bf{IRAS 08572+3915}}&15 April 2004&BPU-2MASS&3x14&3x14&3x14&3x14&6x30&4x60\\
{\bf{IRAS 15250+3609}}&04 March 2004&BPU-2MASS&3x14&3x14&2x30&2x30&6x30&4x60\\
{\bf{UGC 5101}}&15 November 2003$^{a}$&BPU-2MASS&3x14&3x14&2x30&2x30&6x30&4x60\\
{\bf{IRAS 22491-1808}}&24 June 2004&BPU&1x60&1x60&2x30&2x30&2x120&4x60\\
{\bf{IRAS 12112+0305}}&04 January 2004&BPU&3x14&3x14&2x30&2x30&2x120&4x60\\
{\bf{IRAS 14348-1447}}&07 February 2004&BPU&1x60&1x60&2x30&2x30&2x120&4x60\\
\enddata

\tablecomments{The observation date, IRS peak-up type, and integration times for each IRS slit are
given for the 10 BGS ULIRGs.  IRS peak-ups for all sources used the blue array and high accuracy, 
but were performed on offset stars selected via the Staring Mode AOT for the first seven ULIRGs.  In each case, the quantity is cycle ``x" ramp time (in sec) for 
a single nod in staring mode.  There are always two nod positions per IRS slit.
$^{a}$The SL and LL
observations for UGC 5101 were performed on 23 March 2004, while the
SH and LH observations were performed on 15 November 2003.}

\end{deluxetable}

\begin{deluxetable}{lccccc}
\tablewidth{0pc}
\footnotesize
\tabletypesize{\scriptsize}
\tablecaption{Emission Features\label{tbl-1}}
\tablewidth{0pc}
\tablehead{
\colhead{\bf{Line $\mu$m}}&
\colhead{\bf{Mrk 231}}&
\colhead{\bf{Arp 220}}&
\colhead{\bf{05189-2524}}&
\colhead{\bf{Mrk 273}}&
\colhead{\bf{08572+3915}}
}

\startdata
{\bf{PAH 6.2}}&78.6($\pm$29.6)&225($\pm$16)&72.2($\pm$5.1)&149($\pm$22)&$< 36.2$\\
 &10.9($\pm$4.2)&253.3($\pm$17.2)&34.5($\pm$0.8)&171.0($\pm$31.1)&$< 12.3$\\
{\bf{PAH 7.7}}& & & & & \\
 & & & & & \\
{\bf{PAH 8.6}}& & & & & \\
 & & & & & \\
{\bf{H$_{2} $ S(3) 9.665}}&4.33($\pm$1.02)&7.86($\pm0.44$)&3.49($\pm0.63$)&11.70($\pm$0.67)&$< 0.7$\\
 &1.3($\pm$0.2)&79.9($\pm4.5$)&3.2($\pm0.6$)&134.1($\pm$4.5)&$< 25.9$\\
{\bf{[SIV] 10.511}}&$< 3.1$&$< 0.6$&6.93($\pm$0.91)&12.66($\pm$1.27)&$< 0.5$\\
 &$< 1.0$&$< 6.2$&5.4($\pm$0.5)&105.4($\pm$3.9)&$< 9.3$\\
{\bf{PAH 11.3}}&75.57($\pm5.29$)&137.53($\pm9.63$)&78.99($\pm5.53$)&79.26($\pm5.55$)&$< 13.1$\\
 &22.1($\pm1.3$)&520.5($\pm8.9$)&58.2($\pm2.5$)&295.9($\pm4.5$)&$< 50.2$\\
{\bf{H$_{2} $ S(2) 12.279}}&3.93($\pm$1.42)&10.33($\pm$0.43)&1.20($\pm$0.68)&6.31($\pm$0.50)&0.61($\pm$0.15)\\
 &0.9($\pm$0.2)&10.3($\pm$0.2)&0.7($\pm$0.2)&10.5($\pm0.8$)&0.6($\pm$0.1)\\
{\bf{PAH 12.7}}& &306.96($\pm8.77$)& &118.77($\pm5.90$)\\
 & &237.7($\pm6.8$)& &176.2($\pm8.8$)\\
{\bf{[NeII] 12.81}}4&21.11($\pm$1.74)&67.79($\pm$3.64)&21.64($\pm$2.53)&53.48($\pm$1.90)&8.36($\pm$0.69)\\
 &4.0($\pm$0.2)&40.8($\pm$0.3)&12.0($\pm$0.1)&65.5($\pm$1.7)&6.1($\pm$0.4)\\
{\bf{C$_{2}$H$_{2}$ 13.7}}&8.7($\pm1.2$)&7.0($\pm0.8$)& & &5.9($\pm0.8$)\\
 &1.7($\pm0.3$)&4.6($\pm0.3$)& & &4.5($\pm0.6$)\\
{\bf{HCN 14.0}}& &5.7($\pm0.5$)& & &1.8($\pm0.5$)\\
 & &3.6($\pm0.4$)& & &1.4($\pm0.2$)\\
{\bf{[NeV] 14.322}}&$< 2.9$&$< 1.6$&18.36($\pm$3.42)&12.86($\pm$1.41)&$< 1.3$\\
 &$<0.6$&$<1.0$&9.0($\pm$0.3)&22.6($\pm$0.8)&$< 1.0$\\
{\bf{[ClII] 14.368}}&$< 2.9$&2.55($\pm$0.53)&1.34($\pm$0.55)&1.95($\pm$0.58)&$< 1.3$\\
 &$<0.6$&1.5($\pm$0.1)&0.8($\pm$0.1)&21.1($\pm$0.4)&$< 1.0$\\
{\bf{[NeIII] 15.555}}&4.65($\pm$1.72)&8.69($\pm$0.38)&18.59($\pm$1.55)&42.71($\pm$3.42)&2.46($\pm$0.50)\\
 &1.0($\pm$0.4)&6.4($\pm$0.3)&12.6($\pm$0.3)&71.7($\pm$5.5)&2.8($\pm$0.5)\\
{\bf{PAH 16.4}}& &9.74($\pm0.82$)& &5.89($\pm0.91$)& \\
 & &8.8($\pm0.7$)& &11.0($\pm1.7$)\\
{\bf{H$_{2} $ S(1) 17.035}}&11.42($\pm$2.78)&20.46($\pm$1.05)&3.94($\pm$0.55)&11.74($\pm$0.60)&1.37($\pm$0.23)\\
 &2.8($\pm$0.6)&19.8($\pm$0.8)&2.5($\pm$0.1)&22.5($\pm$0.6)&2.3($\pm$0.6)\\
{\bf{PAH 17.4}}& & & &1.18($\pm0.39$)\\
 & & & &2.3($\pm0.8$)\\
{\bf{[SIII] 18.713}}&$< 5.9$&6.04($\pm$0.48)&4.11($\pm$1.61)&17.51($\pm$1.13)&1.84($\pm$0.51)\\
 &$<1.5$&5.5($\pm$0.3)&2.2($\pm$0.4)&31.4($\pm$0.9)&4.1($\pm$0.2)\\
{\bf{[NeV] 24.318}}&$< 5.6$&$< 5.9$&11.67($\pm$1.76)&17.70($\pm$1.80)&$< 1.9$\\
 &$< 1.2$&$< 1.4$&4.9($\pm$0.4)&13.0($\pm$2.1)&$< 1.9$\\
{\bf{[OIV] 25.890}}&$< 6.7$&$< 8.2$&25.53($\pm$2.48)&63.76($\pm$4.79)&$< 2.1$\\
 &$< 1.4$&$< 1.5$&16.8($\pm$1.7)&41.7($\pm$1.6)&$< 1.6$\\
{\bf{[FeII] 25.988}}& & &6.47($\pm$1.04)&4.65($\pm$0.79)\\
 & & &5.2($\pm$2.3)&25.5($\pm$13.6)\\
{\bf{[SIII] 33.481}}&$< 27.5$&$< 35.9$& &34.86($\pm$6.34)&$< 7.7$\\
 &$< 5.6$&$< 3.6$& &13.9($\pm$3.2)&$< 5.4$\\
{\bf{[SiII] 34.815}}& & &10.38($\pm$1.47)&44.02($\pm$7.46)\\
 & & &4.8($\pm0.7$)&15.4($\pm$0.1)\\
\enddata

\tablecomments{Fine structure, H$_2$ emission lines, and PAH emission features in the BGS ULIRG spectra.  The
flux (in units of $10^{-14}$erg cm$^{-2}$ s$^{-1}$), and the rest-frame equivalent
width (in units of $10^{-3} \mu$m), are given for each line. 
The equivalent widths are listed directly
below the fluxes for each emission line.  In most cases, the fluxes and equivalent widths
of the fine-structure and H$_2$ lines are derived from single Gaussian fits to the
high-resolution IRS data, except for lines blueward of $\sim9.6\mu$m, which are only
measured in the low-resolution spectra.
Uncertainties in the fluxes and equivalent widths are listed in
parentheses, and are the larger of either the Gaussian fit, or the
difference in the two nod positions.  Upper limits ($3\sigma$) are given for undetected, key diagnostic features.}

\end{deluxetable}

\begin{deluxetable}{lccccc}
\tablewidth{0pc}
\footnotesize
\tabletypesize{\scriptsize}
\tablecaption{Emission Features\label{tbl-1}}
\tablewidth{0pc}
\tablehead{
\colhead{\bf{Line $\mu$m}}&
\colhead{\bf{15250+3609}}&
\colhead{\bf{UGC 5101}}&
\colhead{\bf{22491-1808}}&
\colhead{\bf{12112+0305}}&
\colhead{\bf{14348-1447}}
}

\startdata
{\bf{PAH 6.2}}&21.3($\pm$1.5)&168($\pm$18)&44.8($\pm$3.1)&88.6($\pm$6.2)&53.5($\pm$3.8)\\
 &22.5($\pm$0.9)&188.0($\pm$22.9)&594.3($\pm$35.4)&516.7($\pm$27.8)&253.5($\pm$3.9)\\
{\bf{PAH 7.7}}& &378($\pm$7)&99($\pm$7)&159($\pm$4)\\
 & &419($\pm$11)&671($\pm$34)&569($\pm$22)\\
{\bf{PAH 8.6}}& &33.5($\pm$0.9)&11.1($\pm$0.2)&15.4($\pm$0.7)\\
 & &51($\pm$1)&107($\pm$2)&80($\pm$4)\\
{\bf{H$_{2} $ S(3) 9.665}}&0.82($\pm$0.08)&3.28($\pm$0.30)&1.36($\pm$0.39)& &3.58($\pm$0.29)\\
 &47.7($\pm$0.6)&37.7($\pm$3.11)&873.1($\pm$46.2)& &210.9($\pm$75.1)\\
{\bf{[SIV] 10.511}}&$< 0.7$&1.68($\pm$0.50)&0.32($\pm$0.11)&0.71($\pm$0.17)&0.37($\pm$0.14)\\
 &$< 82.7$&14.2($\pm$4.3)&51.7($\pm$24.6)&20.5($\pm$1.2)&11.9($\pm$3.6)\\
{\bf{PAH 11.3}}&27.59($\pm1.93$)&107.12($\pm7.50$)&30.39($\pm2.13$)&48.71($\pm3.41$)&39.92($\pm2.79$)\\
 &385.7($\pm6.9$)&399.8($\pm5.4$)&597.4($\pm18.5$)&610($\pm11.7$)&673.7($\pm18.5$)\\
{\bf{H$_{2} $ S(2) 12.279}}&0.69($\pm$0.23)&3.08($\pm$0.57)&0.92($\pm$0.16)&2.35($\pm$0.22)&2.48($\pm$0.18)\\
 &2.4($\pm$0.4)&6.4($\pm$1.2)&11.6($\pm$0.6)&17.8($\pm$0.7)&24.0($\pm$0.3)\\
{\bf{PAH 12.7}}&41.59($\pm1.45$)&86.40($\pm4.86$)&28.55($\pm1.16$)&35.41($\pm2.11$)&27.82($\pm1.14$) \\
 &92.6($\pm3.2$)&174.6($\pm9.8$)&277.3($\pm11.3$)&282.2($\pm16.7$)&246.9($\pm10.1$)\\
{\bf{[NeII] 12.814}}&11.89($\pm$0.93)&43.56($\pm$2.37)&6.97($\pm$0.68)&19.59($\pm$2.94)&13.82($\pm$1.53)\\
 &24.9($\pm$0.4)&66.8($\pm$0.9)&53.2($\pm$1.9)&102.6($\pm$15.4)&87.7($\pm$02.4)\\
{\bf{C$_{2}$H$_{2}$ 13.7}}&5.6($\pm0.7$)\\
 &12.0($\pm0.7$)\\
{\bf{HCN 14.0}}&3.2($\pm0.4$)\\
 &7.0($\pm0.2$)\\
{\bf{[NeV] 14.322}}&$< 0.9$&5.10($\pm$0.97)&$< 0.3$&$< 0.3$&$< 0.2$\\
 &$< 1.7$&18.1($\pm$0.8)&$< 2.4$&$< 2.1$&$< 1.8$\\
{\bf{[ClII] 14.368}}&$< 0.8$&1.94($\pm$0.04)&$< 0.3$&0.21($\pm$0.05)&$< 0.2$\\
 &$< 1.7$&14.8($\pm$0.5)&$< 2.4$&1.5($\pm$0.4)&$< 1.8$\\
{\bf{[NeIII] 15.555}}&3.18($\pm$0.54)&18.69($\pm$2.19)&2.74($\pm$0.32)&5.39($\pm$0.25)&3.24($\pm$0.28)\\
 &8.6($\pm$0.5)&52.6($\pm$0.9)&23.7($\pm$4.1)&44.8($\pm$1.6)&26.4($\pm$2.6)\\
{\bf{PAH 16.4}}& &7.21($\pm0.68$)&2.19($\pm0.45$)&3.06($\pm0.35$)&3.17($\pm0.50$)\\
 & &19.4($\pm1.8$)&14.8($\pm3.0$)&23.6($\pm2.7$)&20.6($\pm3.2$)\\
{\bf{H$_{2} $ S(1) 17.035}}&1.84($\pm$0.28)&5.83($\pm$0.51)&2.22($\pm$0.21)&5.55($\pm$0.24)&6.45($\pm$0.43)\\
 &5.4($\pm$0.6)&15.4($\pm$0.1)&18.9($\pm$1.5)&39.7($\pm$0.2)&50.7($\pm$0.4)\\
{\bf{PAH 17.4}}& &2.37($\pm0.49$)\\
 & &5.8($\pm1.2$)\\
{\bf{[SIII] 18.713}}&1.50($\pm$0.14)&6.94($\pm$0.77)&3.14($\pm$0.49)&6.16($\pm$0.78)&3.53($\pm$0.32)\\
 &5.2($\pm$0.2)&17.9($\pm$0.4)&30.9($\pm$0.3)&67.2($\pm$11.5)&29.8($\pm$0.1)\\
{\bf{[NeV] 24.318}}&$< 0.9$&3.07($\pm$1.25)&$< 0.8$&$< 0.7$&$< 0.9$\\
 &$< 1.0$&5.7($\pm$1.8)&$< 2.0$&$< 2.2$&$< 2.6$\\
{\bf{[OIV] 25.890}}&$< 1.5$&9.05($\pm$0.78)&$< 1.1$&$< 1.0$&$< 1.2$\\
 &$< 1.4$&21.5($\pm$1.9)&$< 2.2$&$< 2.2$&$< 2.7$\\
{\bf{[FeII] 25.988}}& & &\\
 & & &\\
{\bf{[SIII] 33.481}}&$< 9.4$&19.02($\pm$4.00)&4.82($\pm$1.94)&12.73($\pm$3.82)&6.12($\pm$2.18)\\
 &$< 6.8$&18.1($\pm$0.4)&5.9($\pm$0.7)&12.0($\pm$5.1)&7.9($\pm$6.2)\\
{\bf{[SiII] 34.815}}& &33.57($\pm$4.76)\\
 & &33.1(1.0)\\
\enddata

\tablecomments{Spectral features in the remaining five BGS ULIRG spectra. All quantities
are given as in the previous table.}

\end{deluxetable}

\begin{deluxetable}{lccc}
  \tabletypesize{\scriptsize}
  \tablecaption{Absorption Features}
  \tablewidth{0pc}
  \tablehead{
    \colhead{Object} &
    \colhead{$\tau_{9.7}$}   &
    \colhead{C$_{2}$H$_{2}$} &
    \colhead{HCN}
}
\startdata
{\bf{Mrk 231}}         &0.8($\pm 0.1$)&1.7($\pm 0.3$)&  \\
{\bf{Arp 220}}         &3.3($\pm 0.2$)&4.6($\pm 0.3$)&3.6($\pm 0.4$)\\
{\bf{IRAS 05189-2524}} &0.4($\pm0.1$)&  &  \\
{\bf{Mrk 273}}         &1.8($\pm 0.4$)& &  \\
{\bf{IRAS 08572+3915}} &4.2($\pm 0.1$)&4.5($\pm 0.6$)&1.4($\pm 0.2$)\\
{\bf{IRAS 15250+3609}} &3.8($\pm 0.2$)&12.0($\pm 0.7$)&7.0($\pm 0.2$)\\
{\bf{UGC 5101}}        &1.6($\pm 0.3$)& &  \\
{\bf{IRAS 22491-1808}} &1.1($\pm 0.2$)& &  \\
{\bf{IRAS 12112+0305}} &1.3($\pm 0.3$)& &  \\
{\bf{IRAS 14348-1445}} &1.6($\pm 0.3$)& &  \\

\enddata

\tablecomments{Absorption features in the BGS ULIRG data.  Column 2 is the silicate
optical depth, and columns 3 \& 4 are the EQW (in units of $10^{-3}\mu$m) 
of the C$_{2}$H$_{2}$ and HCN features
at $13.7\mu$m and $14.0\mu$m, respectively.}

\end{deluxetable}

\begin{deluxetable}{lcccccccc}
\tabletypesize{\scriptsize}
\tablecaption{SED Fitting\label{tbl-1}}
\tablewidth{0pc}
\tablehead{
\colhead{\bf{Object}} &
\colhead{\bf{T$_{C}$}} &
\colhead{\bf{T$_{W}$}} &
\colhead{\bf{T$_{H}$}} &
\colhead{\bf{$\tau_{W}$}} &
\colhead{\bf{$\tau_{H}$}} &
\colhead{\bf{L$_{H}$/L$_{tot}$}} &
\colhead{\bf{L$_{1-40}$/L$_{tot}$}} &
\colhead{\bf{L$_{PAH}$/L$_{tot}$}} \\
\colhead{}&
\colhead{(K)}&
\colhead{(K)}&
\colhead{(K)}&
\colhead{}&
\colhead{}&
\colhead{($10^{-2}$)}&
\colhead{($10^{-2}$)}&
\colhead{($10^{-2}$)}
}

\startdata
{\bf{Mrk 231}}&38.1($\pm 0.3$)&151($\pm 5$)&486($\pm 3$)&4.3($\pm 0.1$)&1.3($\pm 0.1$)&13(26)&52(99)&0.6\\
{\bf{Arp 220}}&31.4($\pm 0.1$)&96($\pm 2$)&873($\pm 32$)&10.6($\pm 0.1$)&1.3($\pm 0.3$)&0.5(1)&19(99)&0.4\\
{\bf{05189-2524}}&39($\pm 1$)&143($\pm 1$)&579($\pm 4$)&3.8($\pm 0.1$)&0.6($\pm 0.1$)&12(18)&51(85)&0.8\\
{\bf{Mrk 273}}&36.9($\pm 0.1$)&186($\pm 1$)& &6.5($\pm 0.1$)& & &27(56)&1.2\\
{\bf{08572+3915}}&41.7($\pm 0.3$)&145($\pm 15$)&461($\pm 27$)&10.4($\pm 0.8$)&6.1($\pm 0.3$)&23(263*)&56(274*)&$<0.1$\\
{\bf{15250+3609}}&41.7($\pm 0.2$)&212($\pm 1$)& &7.1($\pm 0.2$)& & &36(86)&1.2\\
{\bf{UGC 5101}}&31.5($\pm 0.1$)&115($\pm 9$)&860($\pm 3$)&5.0($\pm 0.1$)&2.7($\pm 0.1$)&8(41)&27(52)&2.1\\
{\bf{22491-1808}}&37.8($\pm 0.3$)&103($\pm 4$)& &9.3($\pm 0.4$)& & &32(107)&1.8\\
{\bf{12112+0305}}&35.8($\pm 0.1$)&187($\pm 8$)& &5.8($\pm 0.2$)& & &21(37)&1.5\\
{\bf{14348-1447}}&36.6($\pm 0.1$)&177($\pm 2$)& &7.4($\pm 0.1$)& & &25(55)&1.6\\
 & & & & & & & & \\
{\bf{Mrk 1014}}&38.0($\pm 0.4$)&121($\pm 1$)&453($\pm 25$)&1.7($\pm 0.1$)& &10(10)&57(77)&0.5\\
{\bf{Mrk 463e}}&28.3($\pm 0.5$)&130($\pm 2$)&572($\pm 12$)&2.0($\pm 0.1$)&1.4($\pm 0.1$)&33(85)&81(130)&0.4\\
{\bf{NGC 6240}}&29.4($\pm 0.3$)&92($\pm 2$)&704($\pm 20$)&5.2($\pm 0.1$)& &3(3)&31(75)&2.2\\
{\bf{NGC 7714}}&30.2($\pm 0.5$)&84.5($\pm 0.8$)& &0.6($\pm 0.1$)& & &37(42)&4.4\\
\enddata

\tablecomments{Quantities derived from the SED fitting of the BGS ULIRGs and other comparison sources.
Columns 1-3: The temperatures of the cold (C), warm (W)
and hot (H) dust components, in degrees K.  Columns 4-5:  The (screen) absorption ($\tau$)
for the warm and hot dust components.  For the temperatures and absorpbing screens, 
the formal $1\sigma$ uncertainties in the fit are given in parentheses.
Column 7: Ratio of the hot dust luminosity to the total
dust luminosity, in units of $10^{-2}$.
An empty column means there is no evidence for
hot dust with T $>300$K (see discussion in text).
Column 8: Ratio of the dust luminosity from $1-40\mu$m to the total
dust luminosity, in units of $10^{-2}$.   
For columns 7 and 8, luminosity ratios are given both before and after (in parentheses)
a correction for the derived extinction.  *The corrected values for IRAS 08572+3915 are 
overestimated due to a poor fit to the silicate absorption profile (see text for a discussion).
Column 9: Ratio of the total PAH to total dust luminosity, 
in units of $10^{-2}$, uncorrected for extinction. 
The fits for Mrk 1014, Mrk 463e, and NGC 7714 are taken from 
Marshall et al.(2006), while that of NGC 6240 is taken from Armus et al. (2006).}

\end{deluxetable}

\begin{deluxetable}{lccccccc}
\tabletypesize{\scriptsize}
\tablecaption{Spectral Classifications\label{tbl-1}}
\tablewidth{0pc}
\tablehead{
\colhead{\bf{Object}} &
\colhead{\bf {opt/NIR}} &
\colhead{\bf{L$_{HX}$/L$_{IR}$}} &
\colhead{\bf {IRS}} &
\colhead{\bf {IRS}} &
\colhead{\bf {IRS}} &
\colhead{\bf {IRS}} &
\colhead{\bf {IRS}} \\
\colhead{}&
\colhead{class}&
\colhead{($10^{-4}$)}&
\colhead{[NeV]/[NeII]}&
\colhead{[OIV/[NeII]}&
\colhead{($6.2\mu$m EQW)} &
\colhead{MIR slope} &
\colhead{L$_{PAH}$/L$_{tot}$} 
}

\startdata
{\bf{Mrk 231}}&S1$^{1}$&$35^{1a}$&$<10$\%&$<10$\%&98\%&$\sim 100$\%\\
{\bf{Arp 220}}&L$^{2}$&$0.2^{2a}$&$<2$\%&$<5$\%&50\%&50\%&90\%\\
{\bf{05189-2524}}&S2/S1$^{6}$&$27^{3a}$&60\%&33\%&94\%&95\%\\
{\bf{Mrk 273}}&S2&$16^{4a}$&17\%&33\%&70\%&75\%&73\%\\
{\bf{08572+3915}}&L$^{2}$& &$<10$\%&$<10$\%&$>95$\%&$>95$\%&$>95$\%\\
{\bf{15250+3609}}&SB&$0.4^{5a}$&$<5$\%&$<5$\%&96\%&$95$\%&73\%\\
{\bf{UGC 5101}}&L$^{3}$&$15^{6a}$&8\%&6\%&65\%&80\%&52\%\\
{\bf{22491-1808}}&SB$^{1,3}$&$0.1^{5a}$&$<5$\%&$<5$\%&$<1$\%&$<10$\%&60\%\\
{\bf{12112+0305}}&L$^{2,4}$&$0.2^{5a}$&$<1$\%&$<5$\%&8\%&30\%&65\%\\
{\bf{14348-1447}}&L$^{1,4}$&$0.3^{5a}$&$<1$\%&$<5$\%&55\%&65\%&64\%\\
 & & & & \\
{\bf{Mrk 1014}}&S1$^{1}$&$94^{7a}$&62\%&50\%&90\%&90\%\\
{\bf{Mrk 463e}}&S2/S1$^{5,6}$&$19^{4a}$&$>99$\%&$>99$\%&$>99$\%&$\sim 100$\%\\
{\bf{NGC 6240}}&L&$250-740^{8a}$&2\%&4\%&15\%&50\%&50\%\\
\enddata

\tablecomments{Spectral classifications of ULIRGs.  Quantities are from the literature or derived from the IRS spectra.
{\bf Column 2}: Optical/NIR 
spectral classifications taken from: 
(1) Sanders et al. 1988,(2) Armus et al. 1989 ,(3) Veilleux et al. 1995,
(4) Kim et al. 1998, (5) Shuder \& Osterbrock 1981, 
(6) Veilleux et al. 1997.  Here, S1 = Seyfert 1, S2 = Seyfert 2, L = LINER, and 
SB = starburst spectra. 
{\bf Column 3}: Ratio of the hard X-ray (2-10 keV) to infrared
luminosity, in units of $10^{-4}$.
The X-ray data are taken from: (1a) Braito et al.
2004, (2a) Iwasawa et al 2005, (3a) Ptak et al. 2003, (4a) Bassani et al.
1999, (5a) Franceschini et al. 2003, (6a) Imanishi et al. 2003, (7a)
Boller et al. 2002, (8a) Vignati et al. 1999, and are all corrected for
extinction (N$_{H}$ as fit to the X-ray data).Seyfert 2 galaxies generally
have values of $50-500$, while Seyfert 1 galaxies have values of $200-2000$ (Ptak et al. 2003).
{\bf Column 4}: The IRS classification (in units of AGN fraction) based upon the [NeV]/[NeII]  
fine-structure emission-line flux ratios.  
{\bf Column 5}: Same as in column 4 but for the [OIV]/[NeII] line flux ratio.
{\bf Column 6}: The IRS classification (in units of AGN fraction) based upon the equivalent width
of the $6.2\mu$m PAH emission feature.
{\bf Column 7}: The IRS classification (in units of AGN fraction) based upon Figure 8.
{\bf Column 8}: The IRS classification (in units of AGN fraction) based upon the ratio of the 
total PAH emission divided by the total dust emission, as derived from the SED fitting
of section 3.4.  Sources used to define the AGN locus (Mrk 231, Mrk 463, Mrk 1014 and 
IRAS 05189-2524) have no entries in this column as they are by this definition, 100\% AGN.
The quantities in columns $4-8$ have not been corrected for extinction.
}

\end{deluxetable}

\end{document}